\documentclass[12pt]{article}

\usepackage{amssymb}
\usepackage{amsmath}
\DeclareMathOperator\artanh{artanh}
\usepackage{siunitx}
\usepackage{geometry}
\usepackage{authblk}
\usepackage{enumitem}
\usepackage{graphicx}
\usepackage[font=footnotesize, labelfont=bf]{caption}
\usepackage{subcaption}
\usepackage{longtable}
\usepackage{multirow}
\usepackage{setspace}
\usepackage{parskip}
\usepackage{rotating}
\usepackage{natbib}
\setcitestyle{round,semicolon}

\title{A Mathematical Model for the Role of Macrophage Chemotactic Emigration in the Early Atherosclerotic Plaque}
\author[1]{Michael G. Watson}
\affil[1]{School of Mathematics and Statistics, University of New South Wales, New South Wales 2052, Australia}
\date{ }

\begin{document}

\maketitle
\abovedisplayskip=12pt
\belowdisplayskip=12pt
\setlength{\jot}{2ex}

\begin{abstract}
Atherosclerotic plaques are fatty, cellular lesions that form in artery walls. The early plaque contains monocyte-derived macrophages, which are recruited to consume locally bound lipid deposits. Plaque progression is characterised by an imbalance in the rates of cell entry and exit from the plaque, which can occur if macrophages die \emph{in situ} rather than leave by emigration. The mechanisms that regulate macrophage emigration are not well understood, but there is evidence that a chemotactic response can guide macrophages out of the plaque towards the artery wall lymphatics.

In this paper, we develop a novel spatial model of the early plaque to study the implications of macrophage chemotactic emigration. Using mathematical analysis and numerical simulations, we investigate how the properties of the chemotactic response contribute to the spatial characteristics and lipid burden of the model plaque. Calculations of macrophage transit times are found to provide a reliable indicator of long-term plaque lipid burden, and also highlight the potential rate-limiting effect of the internal elastic lamina (IEL) on chemotactic emigration. When macrophage emigration is rate-limited by the IEL, we observe non-monotonic cell and lipid profiles that are associated with macrophage accumulation deep in the plaque. The model further predicts that when the chemoattractant penetrates only a short distance into the plaque, the proportion of emigrating macrophages may increase relative to that for a longer-range signal. The theoretical observations in this study can potentially be used to identify evidence of macrophage emigration in data from real atherosclerotic plaques.
\end{abstract}

\section{Introduction}
Atherosclerotic plaques are fatty, cellular lesions that form in artery walls. The rupture of a mature plaque is a common cause of stroke or myocardial infarction \citep{Hans06}. Early plaque formation is characterised by the accumulation of specialised immune cells called macrophages that ingest lipid deposits in the artery wall \citep{Baec19}. These cells can enter and exit the plaque in response to chemical signals \citep{Kang21}. In this paper, we develop a new reaction-diffusion-type model to study how macrophages and lipids distribute spatially in the early plaque, with a novel focus on the role of directed cell movement in macrophage emigration out of the plaque.      

\begin{figure}[h!]
	\centering		
	\includegraphics[height=5cm]{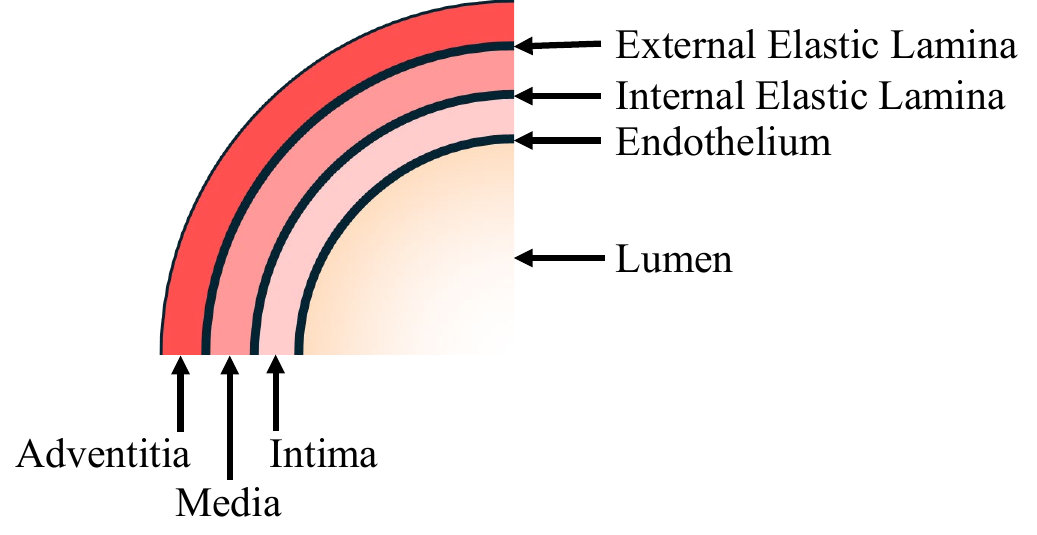}
	\caption{Schematic diagram showing the layered structure of the artery wall (layer widths not to scale). Atherosclerotic plaques form in the \emph{intima}. Blood-derived immune cells enter the plaque from the \emph{lumen} by passing across the \emph{endothelium}, and exit the plaque into the \emph{media} by passing across the \emph{internal elastic lamina}. Chemical signals that direct immune cell emigration out of the plaque are produced in the \emph{adventitia}.}\label{art_schematic}
\end{figure}

Atherosclerotic plaques form in the tunica intima, which is one layer of the multilayered artery wall structure (Figure~\ref{art_schematic}). The intima is located between the endothelium (a thin sheet of endothelial cells that lines the vessel lumen) and the internal elastic lamina (IEL), which is a thin membrane with fenestral holes of a few microns in diameter that facilitate cellular signalling between artery wall layers \citep{Sand09}. Beyond the IEL lies the tunica media and the tunica adventitia. The adventitia contains a system of lymphatic vessels that provide a conduit for fluid, macromolecules, and immune cells to leave the artery wall \citep{Yeo21}.

Plaque formation is initiated by the retention of low-density lipoprotein (LDL) particles in the artery wall. These lipid-bearing particles enter the artery wall from the bloodstream and bind to extracellular matrix proteins in the intima \citep{Bore16}. Retained LDL particles undergo modifications such as oxidation that render them inflammatory \citep{Foge12}. This leads to the localised recruitment of blood-borne monocytes, which enter the plaque via transendothelial migration \citep{Hans06}. Recruited monocytes differentiate into macrophages, which can ingest (phagocytose) lipid from the bound and modified LDL particles. As a plaque matures by the localised accumulation of lipid-filled macrophages, media-derived smooth muscle cells can enter the plaque and form a collagen cap to protect against plaque rupture \citep{Alex12}. In this study, we focus on plaque formation prior to the entry of smooth muscle cells.

Macrophages can undergo apoptosis in the plaque \citep{Taba10}. Any lipid taken up by an apoptotic macrophage prior to its death --- in addition to the endogenous lipid that makes up the cell's membranes and internal structures --- is therefore retained in the intima. Apoptotic macrophages and their internal lipid may be consumed by live macrophages via a process known as efferocytosis \citep{Yin21}. However, if an apoptotic cell is not consumed sufficiently rapidly, it will undergo secondary necrosis \citep{Koji17}. Over time, the accumulation of necrotic cells in a plaque produces a dangerous necrotic core of lipid and cellular debris.

For a plaque to resolve or regress, it is necessary to remove the retained lipid from the plaque that contributes to ongoing inflammation. There are two main mechanisms by which lipid-bearing macrophages can remove lipid from the plaque. The first is reverse cholesterol transport \citep{Yvan10}, whereby plaque-resident macrophages offload ingested lipid to high-density lipoprotein (HDL) particles. The second is macrophage emigration \citep{Llod04}, whereby macrophages migrate out of the plaque carrying their lipid load with them.

Macrophage emigration involves the directed chemotactic migration of macrophages through the IEL and towards the adventitial lymphatics \citep{Moor18}. This chemotactic response is believed to be coordinated by the binding of lymphoid organ-produced C-C chemokine ligand 19 (CCL19) and/or CCL21 to the immune cell receptor C-C chemokine receptor 7 (CCR7) expressed on macrophages \citep{Kang21}. However, macrophage emigration may be inhibited by so-called ``retention factors'' such as netrin-1 and semaphorin-3E that block the macrophage emigratory response \citep{vanG12,Wans13}. These retention factors can be expressed by plaque macrophages in response to factors such as inflammation, lipid loading and hypoxia \citep{Moor18,Kang21}. Macrophage emigration out of plaques is difficult to detect \emph{in vivo}, but progress has been made using bead labelling techniques in mouse models \citep{Feig10,Feig11,vanG12}. Observations suggest that macrophages are more likely to emigrate from a regressing plaque than from a progressing plaque \citep{Rand14}.

The available experimental evidence suggests that the ability of macrophages to exit a plaque via emigration is regulated by complex signalling events that may vary both spatially and temporally within a given plaque. To avoid developing a very complicated model, in this study we propose a simplified mathematical representation of macrophage emigration in which plaque cells migrate chemotactically in response to a generic, externally sourced chemoattractant. We consider a range of different parameter combinations to capture the variation that may occur in the spatial profile of the diffusbile chemoattractant, the levels of chemoattractant receptor expression on macrophages, and the presence of chemotaxis-inhibiting retention factors. In addition, we consider the impact of variability in the capacity of macrophages to navigate their exit from the plaque via the porous IEL.

The process by which macrophages exit a plaque via the IEL in response to an externally sourced chemoattractant has a clear analogy with the cell migration dynamics that may be observed in a Boyden chamber (transwell migration) assay \citep{Kram13}. In this two-chamber assay, a cell population initially resident in the upper chamber is stimulated to migrate through a filter towards a diffusible chemoattractant source in the lower chamber. Building upon the chemotaxis theory first proposed by \citet{Kell71}, several models have been developed to explain the observed dynamics in such assays \citep{Sher93,Chun09,DiCo16}. The current work shares some similarities with these models, but there are also several key differences. In particular, we model a continuous influx of cells rather than a fixed initial population, and we consider the trafficking of lipid cargo by the migrating phagocytic cells.

Mathematical modelling of atherosclerotic plaque formation has been an area of increasing interest over the past 10 to 15 years \citep{Part16,Avge19,Cai21,McAu22}. In particular, a wide range of models have been developed to study the inflammatory response of macrophages to modified LDL in the early plaque. These models include spatially-averaged approaches formulated using either ODEs \citep{Bule12,Cohe14,Isla16, Thon18,Cham23,Cham24b}, or PDEs in which macrophage internalised lipid load is included as a continuous structure variable \citep{Ford19a,Meun19,Wats23}. Spatial studies of the mechanisms of early plaque formation have used reaction-diffusion-type models \citep{ElKh07,Fok12b,Chal17,Thon19,Cham25}, multiphase models \citep{Ahme23,Ahme24}, biomechanical models \citep{Yang16,Mirz20,Silv20,An25}, and agent-based models \citep{Bhui17}.

The first model to explicitly consider the emigration of plaque macrophages was developed by \citet{Ford19a}. Results and analysis from this model suggest that macrophage emigration can provide critical protection against plaque progression. Macrophage emigration has since been included in several other plaque formation models. In these models, macrophage emigration has been assumed to occur at a constant per capita rate \citep{Ford19a,Cham23,Cham24b}, at a variable per capita rate dependent on macrophage lipid load \citep{Wats23}, or, in spatial models, as a passive process in cells proximal to the IEL \citep{Ahme23,Cham25}. To the best of our knowledge, the model proposed in the current study is the first to consider the implications of plaque macrophage emigration as an active process regulated by an explicit chemotactic mechanism.

The remainder of this paper is structured as follows. We present the mathematical model, including a detailed summary of the modelling assumptions, in Section 2. Results from the model are then presented in Section 3. We first consider a reduced version of the model that is amenable to analysis at steady state, and we then perform a numerical investigation of the full nonlinear model. In each case, we consider how macrophage transit times and plaque spatial structure are influenced by the properties of macrophage chemotactic emigration. We discuss our findings in Section 4, and provide concluding remarks in Section 5.

\section{Model Formulation and Parameterisation}
The aim of the model is to investigate the spatial distribution of macrophages and lipids (both intracellular and extracellular) in the early murine plaque, with a particular focus on the role of macrophage emigration. Mice are the most commonly used animal model in experimental studies of atherosclerotic plaque formation \citep{Getz12}. We model plaque formation on the one-dimensional domain $x\in[0,L]$, which represents a cross-section through the annular intima bounded by the endothelium at $x=0$ and by the IEL at $x=L$. This region is assumed to be far from the edges of the plaque so that any variation in the circumferential direction is negligible.

We assume that plaque formation is initiated at time $t=0$ by the presence of retained (and modified) LDL in the intima. We do not explicitly model the transport of free (blood-borne) LDL in the intima, nor the LDL retention process itself. Rather we assume that the intima has a uniform capacity (per unit length) for LDL retention, and that the retained LDL concentration attains this capacity for all $t\geqslant0$. In other words, we assume that the free LDL concentration throughout the domain is sufficiently large, and its retention rate sufficiently rapid, that any retained lipid consumed by macrophages is always instantaneously replenished. A detailed model of LDL transport and retention in the human intima, where spatially non-uniform LDL retention capacity makes an important contribution to plaque spatial structure, was developed in \citet{Cham25}. The current assumptions are considered to provide a reasonable approximation for LDL dynamics in the murine plaque, where the intima is considerably narrower and blood LDL concentrations can be very high in experimental settings.

We assume that the presence of modified LDL in the intima stimulates the localised recruitment of blood-borne monocytes, which rapidly differentiate into macrophages in the plaque. We denote the macrophage density at position $x$ at time $t$ by $M(x,t)$. The corresponding density of internal lipid cargo in these cells is denoted $A_M(x,t)$. The lipid cargo of each macrophage is comprised of: (1) an endogenous lipid quantity $a_0$ that forms the cell's membranes and structures, and (2) any lipid that the cell internalises whilst in the plaque. Two assumed sources of internalised lipid are the lipid contained in macrophages that have undergone apoptosis, and the lipid contained in apoptotic macrophages that have undergone secondary necrosis. We denote the densities of apoptotic and necrotic lipid by $A_P(x,t)$ and $N(x,t)$, respectively. Apoptotic and necrotic lipid are assumed to colocalise with apoptotic and necrotic cells, but we do not explicitly track the densities of these cell types.

\begin{figure}[h!]
	\centering		
	\includegraphics[height=10cm]{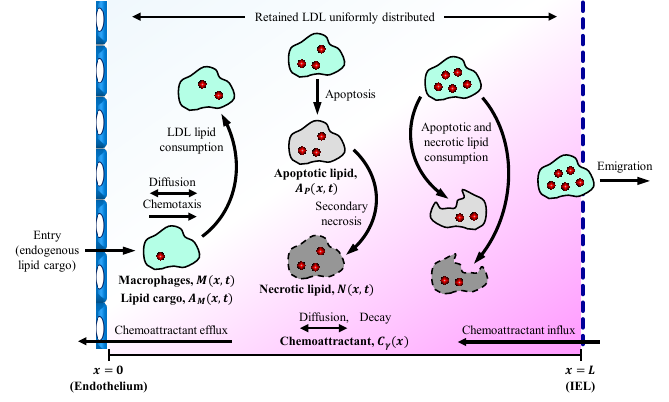}
	\caption{Schematic diagram summarising the key features of the mathematical model. The model domain $x\in[0,L]$ represents the intima, which is bounded by the endothelium and the IEL. The intima contains a steady gradient of macrophage chemoattractant, and a steady concentration of modified LDL that is uniformly distributed in space. Macrophages enter the intima via the endothelium and move randomly with a bias towards increasing chemoattractant concentration. Macrophages can undergo apoptosis or leave the intima via the IEL. Macrophages ingest lipid cargo from modified LDL, from apoptotic cells, and from apoptotic cells that have undergone secondary necrosis.}\label{model_schematic}
\end{figure}

We assume that the emigration of live macrophages out of the plaque across the IEL is mediated by a chemoattractant that is produced outside the intima ($x>L$) and proximal to the adventitial lymphatics. We assume that the transport of this chemoattractant is rapid relative to the transport of macrophages, and we therefore make a quasi-steady state assumption so that the chemoattractant concentration is steady in time. The chemoattractant concentration in the plaque is denoted $C_\gamma(x)$. A schematic diagram of the model setup is provided in Figure~\ref{model_schematic}.

At time $t=0$, we assume that the intima is devoid of live and dead macrophages. We therefore assign the initial conditions:
\begin{equation}
	M(x,0)=A_M(x,0)=A_P(x,0)=N(x,0)=0. \label{dim_init}
\end{equation} 
The differential equations and boundary conditions that govern the macrophage and lipid dynamics for $t>0$ are presented below.

\subsection{Macrophages and Lipid Cargo}
We assume that the macrophages in the plaque are non-interacting and that their movement has both random and chemotactic components. The macrophages can also undergo apoptosis. Assuming that all macrophage behaviours are independent of the internalised lipid load, the PDE that governs the plaque macrophage density $M(x,t)$ is: 
\begin{equation}
	\frac{\partial M}{\partial t} \, = \, D_M \, \frac{\partial^2 M}{\partial x^2} \, - \, \chi_\gamma \, \frac{\partial}{\partial x} \, \Big( \, M \, \frac{dC_\gamma}{dx} \, \Big) \, - \, \beta M \, , \label{dim_M_eqn}
\end{equation}
where $D_M$ is the macrophage diffusion coefficient, $\chi_\gamma$ is the chemotactic coefficient that quantifies the strength of the macrophage flux up the chemoattractant gradient, and $\beta$ is the macrophage apoptosis rate. We remark here that the macrophage transport coefficients $D_M$ and $\chi_\gamma$ are assumed to be constants, and do not depend upon the chemoattractant concentration, as is often considered \citep{Pain19}. This is to avoid the introduction of further unknown parameters, particularly when typical concentrations of the chemoattractants CCL19 and CCL21 in murine plaques are not known.

The entry and the exit of plaque macrophages are governed by boundary conditions. We assume that macrophages are sourced from rapidly differentiating monocytes that enter the plaque across the endothelium in response to inflammatory signalling. Assuming a steady level of inflammation in the plaque, we therefore prescribe a constant macrophage influx $\alpha_M$ at $x=0$ according to the condition:   
\begin{equation}
	\bigg[ \, -\, D_M \, \frac{\partial M}{\partial x} \, + \, \chi_\gamma \, M \, \frac{dC_\gamma}{dx} \, \bigg] \, \bigg|_{\, x \, = \, 0} \, = \, \alpha_M \, . \label{dim_M_BC_0}
\end{equation}
We assume that macrophages exit the plaque and disperse into the media by traversing the IEL. The boundary condition at $x=L$ is therefore given by:
\begin{equation}
	\bigg[ \, -\, D_M \, \frac{\partial M}{\partial x} \, + \, \chi_\gamma \, M \, \frac{dC_\gamma}{dx} \, \bigg] \, \bigg|_{\, x \, = \, L} \, = \, \sigma_M M\big|_{\, x \, = \, L} \, , \label{dim_M_BC_L}
\end{equation}
where the parameter $\sigma_M$ may be interpreted as either the permeability of the IEL to macrophages or the net velocity at which macrophages exit the domain.

Live plaque macrophages are assumed to consume lipid from modified LDL particles at rate $\lambda$ (lipid per cell per unit time), from apoptotic cells at rate $\eta$ (per cell density per unit time), and from necrotic cells at rate $\theta$ (per cell density per unit time). The PDE that governs the internalised lipid density $A_M(x,t)$ in plaque macrophages is therefore given by: 
\begin{equation}
	\frac{\partial A_M}{\partial t} \, = \, D_M \, \frac{\partial^2 A_M}{\partial x^2} \, - \chi_\gamma \, \frac{\partial}{\partial x} \, \Big( \, A_M \, \frac{dC_\gamma}{dx} \, \Big) \, 
	+ \, \big[ \, \lambda \, + \, \eta A_P \, + \, \theta N \, \big] \, M - \, \beta A_M \, , \label{dim_AM_eqn}
\end{equation}
where macrophage apoptosis is seen to reduce the lipid cargo in the live cells, and the spatial flux of internalised lipid has random and chemotactic components with the same transport coefficients as the live cells. This spatial flux term can be formally derived by considering the transport of lipid by macrophages that perform a one-dimensional biased random walk. Note that the presence of the diffusion term allows the internalised lipid to spread spatially, even in the absence of a macroscale macrophage gradient, due to the assumed random movement of macrophages at the microscale.

The boundary conditions for equation \eqref{dim_AM_eqn} follow naturally from those defined above for equation \eqref{dim_M_eqn}. Newly-recruited macrophages that enter the plaque across the endothelium are assumed to carry only their endogenous lipid. The flux of internalised lipid at $x=0$ is therefore given by the condition:  
\begin{equation}
	\bigg[ \, -\, D_M \, \frac{\partial A_M}{\partial x} \, + \, \chi_\gamma \, A_M \, \frac{dC_\gamma}{dx} \, \bigg] \, \bigg|_{\, x \, = \, 0} \, = \, a_0 \, \alpha_M \, . \label{dim_AM_BC_0}
\end{equation}
Macrophages that emigrate out of the plaque across the IEL are assumed to carry their internalised lipid with them. The flux of internalised lipid at $x=L$ is therefore given by:
\begin{equation}
	\bigg[ \, -\, D_M \, \frac{\partial A_M}{\partial x} \, + \, \chi_\gamma \, A_M \, \frac{dC_\gamma}{dx} \, \bigg] \, \bigg|_{\, x \, = \, L} \, = \, \sigma_M A_M\big|_{\, x \, = \, L} \, . \label{dim_AM_BC_L}
\end{equation}

\subsection{Apoptotic and Necrotic Lipid}
As cells that have undergone apoptosis or necrosis are no longer actively motile, we assume that any net movement of apoptotic or necrotic lipid occurs on a much slower timescale than the timescale of interest. We therefore neglect any spatial transport terms in the apoptotic and necrotic lipid equations. Apoptotic lipid accumulates in the plaque from the internalised lipid of macrophages that have undergone apoptosis. As shown in equation \eqref{dim_AM_eqn}, live macrophages consume this apoptotic lipid via efferocytosis at rate $\eta$. However, uncleared apoptotic lipid is assumed to become necrotic lipid at rate $\nu$ (per time) due to secondary necrosis of apoptotic cells. The equation that governs the apoptotic lipid density $A_P(x,t)$ is therefore given by: 
\begin{equation}
	\frac{\partial A_P}{\partial t} \, = \, \beta A_M \, - \, \eta M A_P \, - \, \nu A_P \, . \label{dim_AP_eqn}
\end{equation}
Necrotic lipid can also be consumed by live macrophages via efferocytosis at rate $\theta$. The equation that governs the necrotic lipid density $N(x,t)$ is therefore: 
\begin{equation}
	\frac{\partial N}{\partial t} \, = \, \nu A_P \, - \, \theta M N \, . \label{dim_N_eqn}
\end{equation}

\subsection{Emigratory Chemoattractant}
We assume that the emigratory chemoattractant moves by diffusion and undergoes decay in the plaque. Uptake of the chemoattractant by live macrophages is assumed to be negligible. The equation that governs the steady chemoattractant concentration $C_\gamma(x)$ is therefore:
\begin{equation}
	D_\gamma \, \frac{d^2 C_\gamma}{dx^2} \, = \, \mu_\gamma C_\gamma \, , \label{dim_C_eqn}
\end{equation}
where $D_\gamma$ and $\mu_\gamma$ denote the chemoattractant diffusion coefficient and decay rate in the intima, respectively. We assume that the chemoattractant can diffuse across the endothelium and rapidly disperse in the lumen. The boundary condition prescribed at $x=0$ is therefore:
\begin{equation}
	D_\gamma \, \frac{dC_\gamma}{dx} \, \bigg|_{\, x \, = \, 0} \, = \, \sigma_\gamma C_\gamma \, \big|_{\, x \, = \, 0} \, , \label{dim_C_BC_0}
\end{equation}
where $\sigma_\gamma$ denotes the permeability of the endothelium to the chemoattractant. The chemoattractant is assumed to enter the plaque across the IEL from the media. We assume a constant flux of chemoattractant $\alpha_\gamma$ into the system at $x=L$ according to the condition: 
\begin{equation}
	D_\gamma \, \frac{dC_\gamma}{dx} \, \bigg|_{\, x \, = \, L} \, = \, \alpha_\gamma \, . \label{dim_C_BC_L}
\end{equation}

We remark that the boundary value problem defined by equations \eqref{dim_C_eqn}--\eqref{dim_C_BC_L} is analytically solvable. However, we withhold the solution to this problem until the equations are nondimensionalised in the following section.

\subsection{Model Parameterisation}
The aim of this study is to improve current understanding of how an adventitia-derived chemoattractant may facilitate plaque macrophage emigration out of the intima. Given uncertainty around the responsiveness of macrophages to the chemical signal, the capacity of macrophages to navigate the IEL, and the spatial distribution of the diffusible chemoattractant in the intima, we consider the values of the parameters $\chi_\gamma$, $\sigma_M$, and $\mu_\gamma$ to be variable. \emph{In vivo} observations suggest that macrophages in the murine plaque have limited mobility \citep{Will18}. We therefore focus on parameter combinations that give relatively low chemotactic velocities $\chi_\gamma \, \frac{dC_\gamma}{dx}$ (e.g., $<20$ $\mu$m day$^{-1}$).

The chemical gradient in the domain also depends on the parameters $D_\gamma$, $\sigma_\gamma$, and $\alpha_\gamma$. For the purposes of our analysis, we find that it is not necessary to explicitly specify the dimensional values of these parameters. We do, however, consider their values to be fixed throughout the study.

The remaining model parameters are all supplied with fixed values. The rationale for each estimate is as follows:
\begin{itemize}
	\item[$-$] We take $L=$~50~$\mu$m as a typical intimal width in the early murine plaque prior to the entry of smooth muscle cells \citep{Misr18}.
	\item[$-$] Estimates for the rate of macrophage entry into the plaque per unit area of endothelium are in the approximate range $1\times10^{-2}$ to $7\times10^{-2}$~cell~$\mu$m$^{-2}$~day$^{-1}$ \citep{Thon18}. We take the value $5\times10^{-2}$~cell~$\mu$m$^{-2}$~day$^{-1}$, and convert into appropriate units by assuming that the effective 1D domain has a representative transverse cross-sectional area of $10^5$~$\mu$m$^2$. This gives $\alpha_M=$~500~cells~day$^{-1}$.
	\item[$-$] We set $D_M=$~400~$\mu$m$^2$~day$^{-1}$ for the macrophage diffusion coefficient, which is at the lower end of estimates made using \emph{in vitro} experimental data \citep{Owen97}.
	\item[$-$] Existing estimates for the rate of plaque macrophage apoptosis vary from 0.03~day$^{-1}$ for cells without ingested lipid \citep{Cham24b} to 2.4~day$^{-1}$ for cells laden with ingested lipid \citep{Thon18}. Since we do not consider macrophage apoptosis to depend on lipid load in this study, we take the intermediate value $\beta=$~0.2~day$^{-1}$.
	\item[$-$] Observations suggest that cell lysis occurs within 0.5 to 1 day after apoptosis \citep{Coll97}. We therefore set $\nu=1.2$~day$^{-1}$ for the apoptotic cell secondary necrosis rate.
	\item[$-$] We take the macrophage endogenous lipid content to be the representative unit of lipid mass in the model. Correspondingly, we have $a_0=1$~lipid~cell$^{-1}$.
	\item[$-$] We estimate the rate of macrophage lipid consumption from modified LDL to be $\lambda=$~0.1~lipid~cell$^{-1}$~day$^{-1}$.
	\item[$-$] A study of macrophage efferocytosis by \citet{Mare05} predicts a lower bound of approximately $2\times10^5$~$\mu$m$^3$~cell$^{-1}$~day$^{-1}$ for the rate of engulfment of apoptotic cells \emph{in vitro}. Since the \emph{in vivo} efferocytosis rate is likely to be smaller, and since impaired efferocytosis is considered a hallmark of plaque formation \citep{Koji17}, we conservatively estimate the much reduced value $1\times10^3$~$\mu$m$^3$~cell$^{-1}$~day$^{-1}$. Dividing by the assumed transverse cross-sectional area of the domain, we obtain the estimate $\eta=$~$0.01$~$\mu$m~cell$^{-1}$~day$^{-1}$ for the apoptotic lipid consumption rate.
	\item[$-$] The rate of apoptotic cell phagocytosis is believed to be much more efficient than the rate of phagocytosis of necrotic material \citep{Koji17}. We therefore set the value of $\theta$ relative to $\eta$, and choose $\theta=\frac{\eta}{3}$ for the necrotic lipid consumption rate.
\end{itemize}

A concise summary of the dimensional model parameters is provided in Table \ref{dim_params}.

\begin{table}[h!]
	\centering
	\footnotesize
	\renewcommand{\arraystretch}{1.2}
	\begin{tabular}{| l | l | l |}
		\hline
		Parameter & Description & Value \\ \hline
		$L$ & Intimal width & 50 $\mu$m \\ \hline
		$D_M$ & Macrophage diffusion coefficient & 400 $\mu$m$^2$ day$^{-1}$ \\ \hline
		$\chi_\gamma$ & Macrophage chemotaxis coefficient & Variable \\ \hline
		$\beta$ & Macrophage apoptosis rate & 0.2 day$^{-1}$ \\ \hline
		$\alpha_M$ & Macrophage recruitment rate & 500 cells day$^{-1}$ \\ \hline
		$\sigma_M$ & Permeability of IEL to macrophages & Variable \\ \hline
		$\lambda$ & Macrophage modified LDL lipid consumption rate & 0.1 lipid cell$^{-1}$ day$^{-1}$ \\ \hline
		$\eta$ & Macrophage apoptotic lipid consumption rate & 0.01 $\mu$m cell$^{-1}$ day$^{-1}$ \\ \hline
		$\theta$ & Macrophage necrotic lipid consumption rate & $\eta/3$ \\ \hline
		$a_0$ & Macrophage endogenous lipid quantity & 1 lipid cell$^{-1}$ \\ \hline
		$\nu$ & Apoptotic cell secondary necrosis rate & 1.2 day$^{-1}$ \\ \hline
		$D_\gamma$ & Chemoattractant diffusion coefficient in the intima & Fixed (not specified) \\ \hline
		$\mu_\gamma$ & Chemoattractant decay rate in the intima & Variable \\ \hline
		$\sigma_\gamma$ & Permeability of endothelium to chemoattractant & Fixed (not specified) \\ \hline
		$\alpha_\gamma$ & Chemoattractant influx rate & Fixed (not specified) \\ \hline
	\end{tabular}
	\caption{Summary of the dimensional model parameter values.} \label{dim_params}
\end{table}

\subsection{Model Nondimensionalisation}
In this section, we nondimensionalise the system of dimensional equations \eqref{dim_init}--\eqref{dim_C_BC_L}. Using tildes to denote dimensionless quantities, we nondimensionalise time $t$ with respect to the macrophage apoptosis rate $\beta$, and space $x$ with respect to the intimal width $L$ as follows:
\begin{equation}
	\tilde{x} = \frac{x}{L}, \;\; \tilde{t} = \beta t.
\end{equation}
The live macrophage density, the lipid densities, and the chemoattractant concentration are nondimensionalised according to the following respective relationships:
\begin{equation}
	\frac{M\big(x,t\big)}{\tilde{M}\big(\tilde{x},\tilde{t}\big)} = \frac{\alpha_M}{\beta L},\quad
	\frac{A_M\big(x,t\big)}{\tilde{A}_M\big(\tilde{x},\tilde{t}\big)} = \frac{A_P\big(x,t\big)}{\tilde{A}_P\big(\tilde{x},\tilde{t}\big)} = \frac{N\big(x,t\big)}{\tilde{N}\big(\tilde{x},\tilde{t}\big)} = \frac{a_0 \, \alpha_M}{\beta L},\quad		
	\frac{C_\gamma\big(x\big)}{\tilde{C}_\gamma\big(\tilde{x}\big)} = \frac{\alpha_\gamma}{\mu_\gamma L}. \label{nondim}
\end{equation}
The dimensionless model parameters are defined in Table \ref{dimless_params}. Dropping the tildes for notational convenience, the dimensionless equations for the cell and lipid species are:
\begin{gather}
	\frac{\partial M}{\partial t} \, = \, D_M \, \frac{\partial^2 M}{\partial x^2} \, - \, \chi_\gamma \, \frac{\partial}{\partial x} \, \Big( \, M \, \frac{dC_\gamma}{dx} \, \Big) \, - \, M \, , \label{dimless_M_eqn} \\
	\frac{\partial A_M}{\partial t} \, = \, D_M \, \frac{\partial^2 A_M}{\partial x^2} \, - \, \chi_\gamma \, \frac{\partial}{\partial x} \, \Big( \, A_M \, \frac{dC_\gamma}{dx} \, \Big) \, + \, \big[ \, \lambda \, + \, \eta A_P \, + \, \theta N \, \big] \, M - \, A_M \, , \label{dimless_AM_eqn} \\
	\frac{\partial A_P}{\partial t} \, = \, A_M \, - \, \eta M A_P \, - \, \nu A_P \, , \label{dimless_AP_eqn} \\
	\frac{\partial N}{\partial t} \, = \, \nu A_P \, - \, \theta M N \, . \label{dimless_N_eqn}
\end{gather}
These equations are subject to the initial conditions:
\begin{equation}
	M(x,0)=A_M(x,0)=A_P(x,0)=N(x,0)=0, \label{dimless_init}
\end{equation}
and equations (\ref{dimless_M_eqn}) and (\ref{dimless_AM_eqn}) are also subject to the boundary conditions:
\begin{gather}
	\bigg[ \, -\, D_M \, \frac{\partial M}{\partial x} \, + \, \chi_\gamma \, M \, \frac{dC_\gamma}{dx} \, \bigg] \, \bigg|_{\, x \, = \, 0} \, = \, 1 \, , \label{dimless_M_BC_0} \\
	\bigg[ \, -\, D_M \, \frac{\partial M}{\partial x} \, + \, \chi_\gamma \, M \, \frac{dC_\gamma}{dx} \, \bigg] \, \bigg|_{\, x \, = \, 1} \, = \, \sigma_M M\big|_{\, x \, = \, 1} \, , \label{dimless_M_BC_1} \\
	\bigg[ \, -\, D_M \, \frac{\partial A_M}{\partial x} \, + \, \chi_\gamma \, A_M \, \frac{dC_\gamma}{dx} \, \bigg] \, \bigg|_{\, x \, = \, 0} \, = \, 1 \, , \label{dimless_AM_BC_0} \\
	\bigg[ \, -\, D_M \, \frac{\partial A_M}{\partial x} \, + \, \chi_\gamma \, A_M \, \frac{dC_\gamma}{dx} \, \bigg] \, \bigg|_{\, x \, = \, 1} \, = \, \sigma_M A_M\big|_{\, x \, = \, 1} \, . \label{dimless_AM_BC_1}
\end{gather}

The dimensionless equation for the chemoattractant is:
\begin{equation}
	\frac{d^2 C_\gamma}{dx^2} \, = \, \omega^2_\gamma \, C_\gamma \, , \label{dimless_C_eqn}
\end{equation}
with corresponding boundary conditions:
\begin{gather}
	\frac{dC_\gamma}{dx} \, \bigg|_{\, x \, = \, 0} \, = \, \sigma_\gamma C_\gamma \, \big|_{\, x \, = \, 0} \, , \label{dimless_C_BC_0} \\
	\frac{dC_\gamma}{dx} \, \bigg|_{\, x \, = \, 1} \, = \, \omega^2_\gamma \, . \label{dimless_C_BC_1}
\end{gather}
The boundary value problem defined by equations (\ref{dimless_C_eqn})--(\ref{dimless_C_BC_1}) can be solved analytically to give:
\begin{equation}
	C_\gamma\big(x\big) \, = \, \omega_\gamma \, \frac{\omega_\gamma \cosh(\omega_\gamma \, x) + \sigma_\gamma \sinh(\omega_\gamma \, x)}{\omega_\gamma \sinh(\omega_\gamma) + \sigma_\gamma \cosh(\omega_\gamma)} \, ,
\end{equation}
such that:
\begin{equation}
	\frac{dC_\gamma}{dx} \, = \, \omega^2_\gamma \, \frac{\omega_\gamma \sinh(\omega_\gamma \, x) + \sigma_\gamma \cosh(\omega_\gamma \, x)}{\omega_\gamma \sinh(\omega_\gamma) + \sigma_\gamma \cosh(\omega_\gamma)} \, . \label{dimless_dC_dx}
\end{equation}

The complete model is then defined by equations (\ref{dimless_M_eqn})--(\ref{dimless_N_eqn}), subject to the initial conditions (\ref{dimless_init}), the boundary conditions (\ref{dimless_M_BC_0})--(\ref{dimless_AM_BC_1}), and equation (\ref{dimless_dC_dx}) for the chemical concentration gradient. Numerical solutions of this system of equations (see Section~\ref{results_full}) are generated by the method of lines. The spatial derivatives in equations (\ref{dimless_M_eqn}) and (\ref{dimless_AM_eqn}) are approximated with a uniform central differencing scheme, and the resulting system of time-dependent ODEs is solved via the \emph{ode15s} routine in MATLAB (Version R2023a).

\begin{table}[h!]
	\centering
	\footnotesize
	\renewcommand{\arraystretch}{1.5}
	\begin{tabular}{| l | l | l | l |}
		\hline
		Parameter & Definition & Description & Value \\ \hline
		$\tilde{D}_M$ & $\frac{D_M}{\beta L^2}$ & Macrophage diffusion coefficient & 0.8\\ \hline
		$\tilde{\chi}_\gamma$ & $\frac{\chi_\gamma \, \alpha_\gamma}{\mu_\gamma \, \beta L^3}$ & Macrophage chemotaxis coefficient & Variable\\ \hline
		$\tilde{\sigma}_M$ & $\frac{\sigma_M}{\beta L}$ & Permeability of IEL to macrophages & Variable\\ \hline
		$\tilde{\lambda}$ & $\frac{\lambda}{a_0 \, \beta}$ & Macrophage modified LDL lipid consumption rate & 0.5\\ \hline
		$\tilde{\eta}$ & $\frac{\eta \, \alpha_M}{\beta^2 L}$ & Macrophage apoptotic lipid consumption rate & 2.5\\ \hline
		$\tilde{\theta}$ & $\frac{\theta \, \alpha_M}{\beta^2 L}$ & Macrophage necrotic lipid consumption rate & $\tilde{\eta}/3$\\ \hline
		$\tilde{\nu}$ & $\frac{\nu}{\beta}$ & Apoptotic cell secondary necrosis rate & 6\\ \hline
		$\tilde{\omega}_\gamma$ & $L\,\sqrt{\frac{\mu_\gamma}{D_\gamma}}$ & Reciprocal of chemoattractant diffusion distance & Variable\\ \hline
		$\tilde{\sigma}_\gamma$ & $\frac{\sigma_\gamma L}{D_\gamma}$ & Permeability of endothelium to chemoattractant & 0.1\\ \hline
	\end{tabular}
	\caption{Summary of the dimensionless model parameter values.} \label{dimless_params}
\end{table} 

\section{Results}

\subsection{Reduced Model (Spatially Uniform Chemotaxis)}\label{results_reduced}
In this section, we simplify the model by assuming that the emigratory chemoattractant has a spatially linear profile such that $\frac{dC_\gamma}{dx}$ is constant for all $x$. Physically, this represents a scenario where the chemical diffusion distance in the intima is large (i.e., $\omega_\gamma \ll 1$). Specifically, we assume that the chemotactic velocity $\chi_\gamma \, \frac{dC_\gamma}{dx}$ takes the constant value $h>0$ for all $x$. Under this assumption, equations (\ref{dimless_M_eqn}) and (\ref{dimless_AM_eqn}) become linear advection-diffusion equations, whose steady state solutions can be studied in detail. In Section~\ref{results_full}, we will consider the conditions under which this reduced model provides a reasonable approximation to the full model presented above. 

\subsubsection{Macrophage Transit Time}
We first calculate the average time $\tau$ that a macrophage would take to transit through the model plaque at steady state. The transit time is of interest because it correlates with the lipid quantity that a macrophage consumes while in the plaque, as well as the likelihood that a macrophage dies in the plaque before emigrating. Denoting the steady state cell concentration by $M^*\big(x\big)$, we consider the equation:
\begin{equation}
D_M \frac{d^2 M^*}{dx^2} \, - \, h \, \frac{dM^*}{dx} \, = \, 0 \, , \label{SS_M_trans} \\
\end{equation}
which is the steady state form of equation (\ref{dimless_M_eqn}), in which only the transport terms are retained. Solving equation (\ref{SS_M_trans}) subject to the boundary conditions:
\begin{equation}
\bigg[ \, -D_M \, \frac{dM^*}{dx} \, + \, hM^* \, \bigg] \, \bigg|_{\, x \, = \, 0} \, = \, 1 \, , \label{SS_M_BC_0}
\end{equation}
and:
\begin{equation}
\bigg[ \, -D_M \, \frac{dM^*}{dx} \, + \, hM^* \, \bigg] \, \bigg|_{\, x \, = \, 1} \, = \, \sigma_M M^*\big|_{\, x \, = \, 1} \, , \label{SS_M_BC_1}
\end{equation}
gives:
\begin{equation}
M^*(x) \, = \, \frac{1}{h} \, \Bigg( \,  1 \, + \, \frac{D_M \, \big(h-\sigma_M\big)}{h \, \sigma_M} \, \exp\bigg[-\frac{h\,\big(1-x\big)}{D_M}\,\bigg] \, \Bigg).
\end{equation}
Noting that (\ref{SS_M_BC_0}) imposes unit steady state flux, the average transit time is given by:
\begin{equation}
\tau \, = \, \int_0^1 M^*(x) \, dx \, = \, \frac{1}{h} \, \Bigg( \,  1 \, + \, \frac{D_M \, \big(h-\sigma_M\big)}{h \, \sigma_M} \, \bigg( \, 1-\exp\bigg[-\frac{h}{D_M}\bigg] \, \bigg) \, \Bigg). \label{res_time}
\end{equation}

\begin{figure}[h!]
	\centering
	\begin{subfigure}[b]{0.49\textwidth}
		\centering	
		\includegraphics[height=6.6cm]{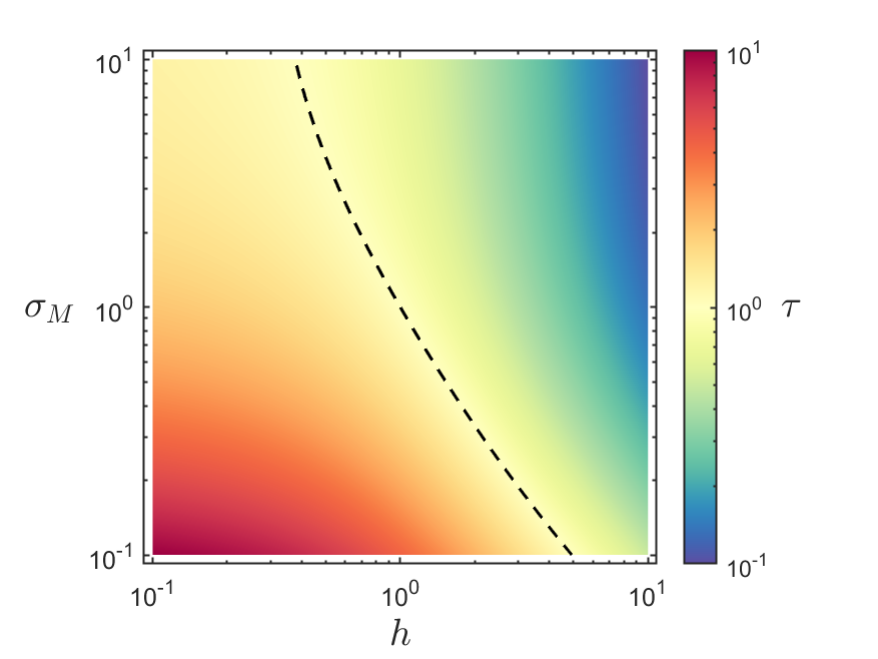}
		\caption{$D_M=0.4$} \label{transit_D_4_10}
	\end{subfigure}
	\begin{subfigure}[b]{0.49\textwidth}
		\centering	
		\includegraphics[height=6.6cm]{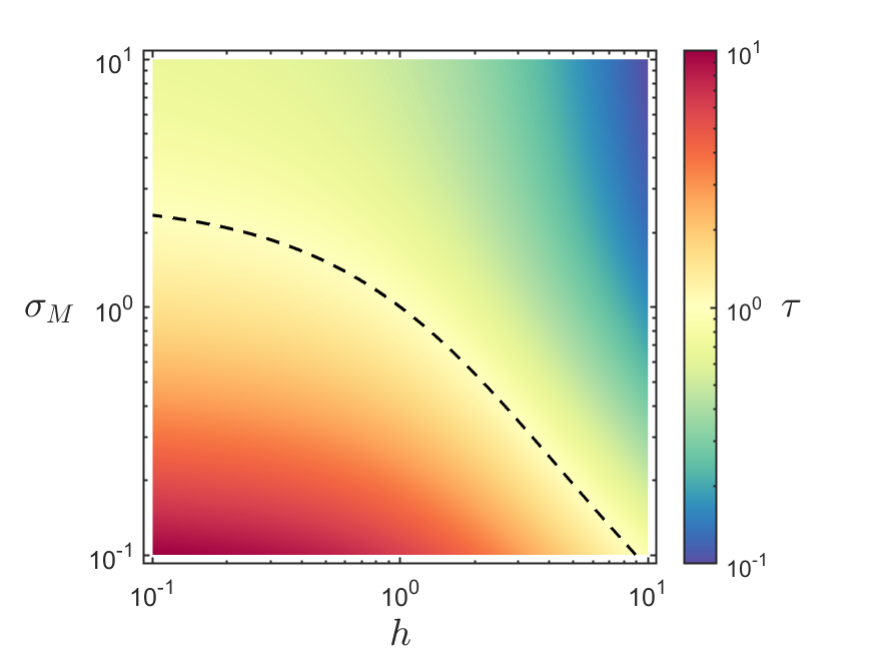}
		\caption{$D_M=0.8$} \label{transit_D_8_10}
	\end{subfigure}
	\caption{Heatmaps of average macrophage transit time $\tau$ given by equation~(\ref{res_time}) for $\left(h,\sigma_M\right)\in\left[0.1,10\right]\times\left[0.1,10\right]$ with (a) $D_M=0.4$, or (b) $D_M=0.8$. The dashed black lines indicate the contours $\tau=1$.
	} \label{transit}
\end{figure}

Transit time distributions as functions of $h$ and $\sigma_M$ are plotted for two different macrophage diffusion coefficients in Figure~\ref{transit}. The dashed black line on each plot indicates the contour on which $\tau = 1$. As time is nondimensionalised with respect to the macrophage apoptosis rate, $\tau=1$ represents the transit time above which a macrophage that enters the plaque is more likely to die than to emigrate. Figure~\ref{transit_D_8_10}, plotted for the base case diffusion coefficient $D_M=0.8$, shows
that the range of $\sigma_M$ values for which a majority of cells can traverse and then exit the plaque before dying increases with increasing $h$. However, the plot also suggests that a majority of cells can exit the plaque by random movement (with minimal directed migration) provided that the IEL is sufficiently permeable. Figure~\ref{transit_D_4_10}, plotted for $D_M=0.4$, demonstrates that this is no longer true if the rate of random movement is sufficiently small. In this case, some level of chemotactic migration appears to be necessary to achieve a transit time $\tau<1$.

In the special case where the flux of cells out of the domain is exactly equal to the chemotactic flux (i.e., $h=\sigma_M$), the transit time is independent of $D_M$ and simplifies to $\tau = \frac{1}{h}$. In this case, macrophages are more likely to die than to emigrate when $h<1$. In dimensional terms, this is when the chemotactic velocity is less than $\beta L =$~10~$\mu$m~day$^{-1}$.

\subsubsection{Steady State Solution}\label{results_reduced_SS}
We now show that an analytical steady state solution can be determined for the system (\ref{dimless_M_eqn})--(\ref{dimless_N_eqn}) in the case of spatially uniform chemotaxis. Again using asterisks to denote steady state variables, the equations for $M^*\big(x\big)$ and $A_M^*\big(x\big)$ can be stated in the form:
\begin{gather}
	D_M \, \frac{d^2 M^*}{dx^2} \, - \, h \, \frac{dM^*}{dx} \, - \, M^* \, = \, 0 \, , \label{SS_M_eqn} \\
	D_M \, \frac{d^2 A_M^*}{dx^2} \, - \, h \, \frac{dA_M^*}{dx} \, + \, \lambda M^* \, = \, 0 \, . \label{SS_AM_eqn}
\end{gather}
Notice that the steady state versions of equations~(\ref{dimless_AP_eqn}) and (\ref{dimless_N_eqn}) have been used to eliminate the parameters $\eta$ and $\theta$ from the equation for $A_M^*$.
Equation (\ref{SS_M_eqn}) is subject to the boundary conditions (\ref{SS_M_BC_0}) and (\ref{SS_M_BC_1}), and equation (\ref{SS_AM_eqn}) is subject to the boundary conditions:
\begin{equation}
\bigg[ \, -D_M \, \frac{dA_M^*}{dx} \, + \, hA_M^* \, \bigg] \, \bigg|_{\, x \, = \, 0} \, = \, 1 \, , \label{SS_AM_BC_0}
\end{equation}
and:
\begin{equation}
\bigg[ \, -D_M \, \frac{dA_M^*}{dx} \, + \, hA_M^* \, \bigg] \, \bigg|_{\, x \, = \, 1} \, = \, \sigma_M A_M^*\big|_{\, x \, = \, 1} \, . \label{SS_AM_BC_1}
\end{equation}

Sequential solution of the above boundary value problems for $M^*$ and $A_M^*$ leads to the following steady state solution for the full PDE system:
\begin{gather}
	M^*\big(x\big) \, = \, M_0 \, \exp\Big( \, \frac{px}{2} \, \Big) \, \bigg[ \, u \, \cosh\Big(u \, \big(1-x\big)\Big) \, + \, \Big( \, s \, - \, \frac{p}{2} \, \Big) \, \sinh\Big(u \, \big(1-x\big)\Big) \, \bigg] \, , \label{M_SS} \\
	A_M^*\big(x\big) \, = \, \frac{r}{p} \, \big( 1 \, + \, \lambda \big) \, \bigg[ \,  1 \, - \, \Big( \, \frac{s - p}{s} \, \Big) \, \exp\Big(-p\big(1-x\big)\Big) \, \bigg] \, - \, \lambda  M^*\big(x\big) \, , \label{AM_SS} \\
	A_P^*\big(x\big) \, = \, \frac{A_M^*\big(x\big)}{\nu \, + \, \eta M^*\big(x\big)} \, , \label{AP_SS} \\
	N^*\big(x\big) \, = \, \frac{\nu A_P^*\big(x\big)}{\theta M^*\big(x\big)} \, , \label{N_SS}
\end{gather}
where:
\begin{equation*}
p \,=\, \frac{h}{D_M}\,, \quad r \,=\, \frac{1}{D_M}\,, \quad s \,=\, \frac{\sigma_M}{D_M}\,,	
\end{equation*}
and:
\begin{equation*}
	M_0 \, = \, \frac{2r}{2su \, \cosh(u) \, + \, \big( sp \, + \, 2r \big) \, \sinh(u)}\,, \quad	
	u \, = \, \frac{\sqrt{p^2+4r}}{2}\,.	
\end{equation*}

We use the solution (\ref{M_SS})--(\ref{N_SS}) to investigate the qualitative solution profiles that are possible for each of the model variables. As the expressions for $A_M^*$, 
$A_P^*$ and $N^*$ are too complicated to analyse directly, we use numerical plots and analysis of the simpler $M^*$ solution to gain insight into the model behaviour. In general, we find that all of the model variables can admit both monotonic and non-monotonic solution profiles at steady state.

\begin{figure}[h!]
	\centering
	\begin{subfigure}[b]{\textwidth}
		\centering	
		\caption{$h=0.3,\,\sigma_M=0.15$}
		\includegraphics[height=3.96cm]{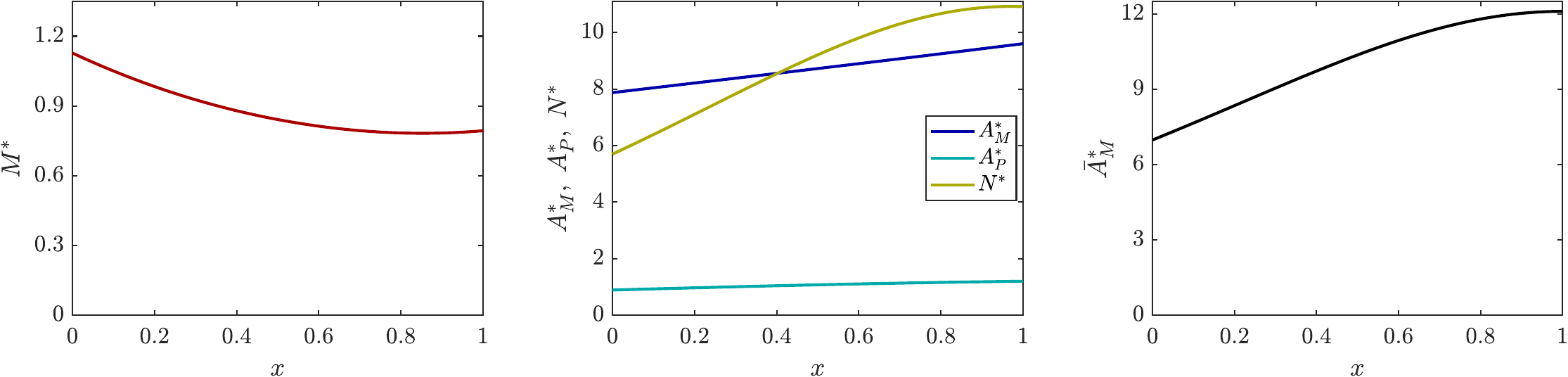}
		 \label{SS_plots_h03_s015}
		 \vspace{-0.4cm}
	\end{subfigure}
	\begin{subfigure}[b]{\textwidth}
		\centering	
		\caption{$h=0.3,\,\sigma_M=0.3$}
		\includegraphics[height=3.96cm]{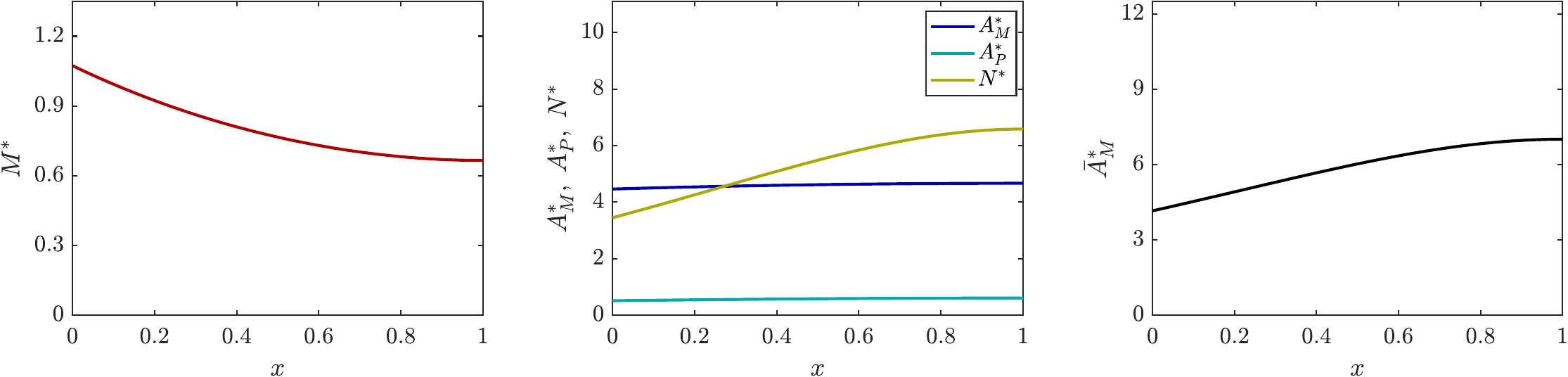}
		 \label{SS_plots_h03_s03}
		 \vspace{-0.4cm}
	\end{subfigure}
	\begin{subfigure}[b]{\textwidth}
		\centering
		\caption{$h=0.3,\,\sigma_M=0.6$}
		\includegraphics[height=3.96cm]{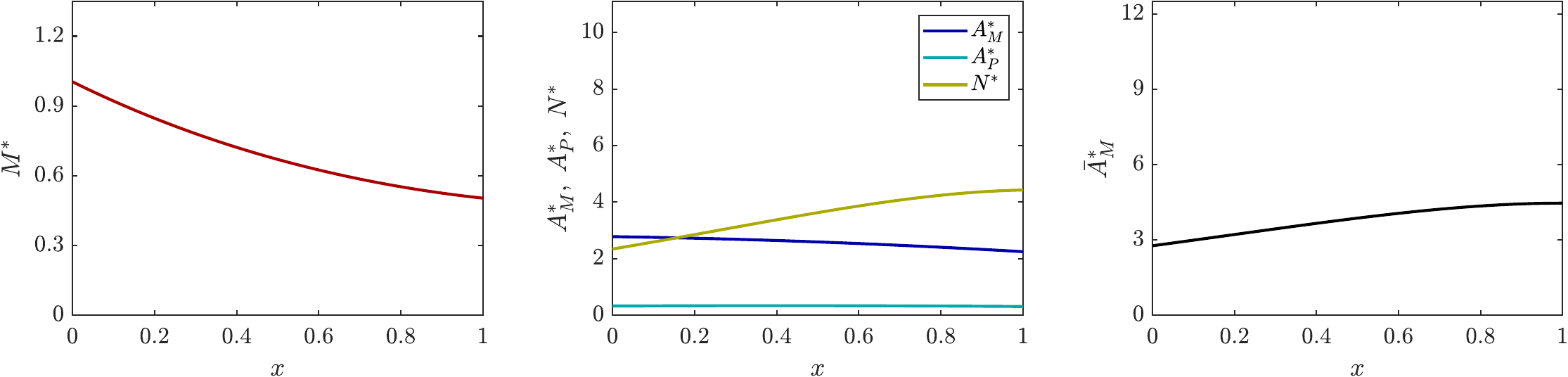}
		 \label{SS_plots_h03_s06}
		 \vspace{-0.4cm}
	\end{subfigure}
	\begin{subfigure}[b]{\textwidth}
		\centering	
		\caption{$h=0.3,\,\sigma_M=1.5$}
		\includegraphics[height=3.96cm]{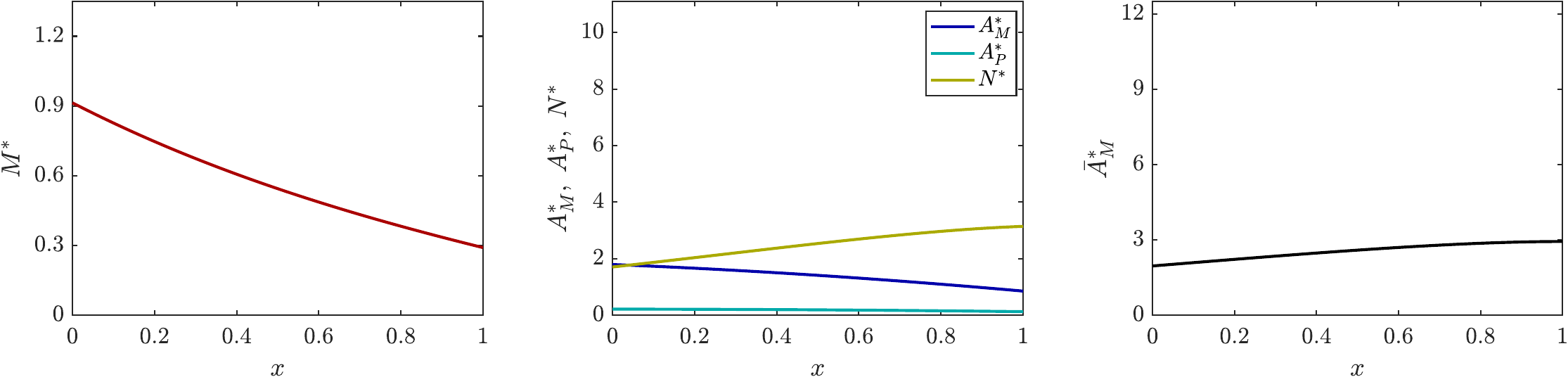}
		 \label{SS_plots_h03_s15}
		 \vspace{-0.6cm}
	\end{subfigure}
	\caption{Steady state solution profiles for $M^*$ (red lines; left), $A_M^*$ (blue lines; middle), $A_P^*$ (cyan lines; middle), $N^*$ (gold lines; middle), and $\bar{A}_M^*=A_M^*/M^*$ (black lines; right) for $h=0.3$ and (a) $\sigma_M=0.15$, (b) $\sigma_M=0.3$, (c) $\sigma_M=0.6$, or (d) $\sigma_M=1.5$. Expressions for $M^*$, $A_M^*$, $A_P^*$, and $N^*$ are given by equations (\ref{M_SS})--(\ref{N_SS}), respectively. All other parameters are set to the values given in Table~\ref{dimless_params}. From equation (\ref{res_time}), the approximate macrophage transit times are (a) $\tau\approx6.11$, (b) $\tau\approx3.33$, (c) $\tau\approx1.94$, and (d) $\tau\approx1.11$.} \label{SS_plots_h03}
\end{figure}

\begin{figure}[h!]
	\centering
	\begin{subfigure}[b]{\textwidth}
		\centering	
		\caption{$h=1.5,\,\sigma_M=0.15$}
		\includegraphics[height=3.96cm]{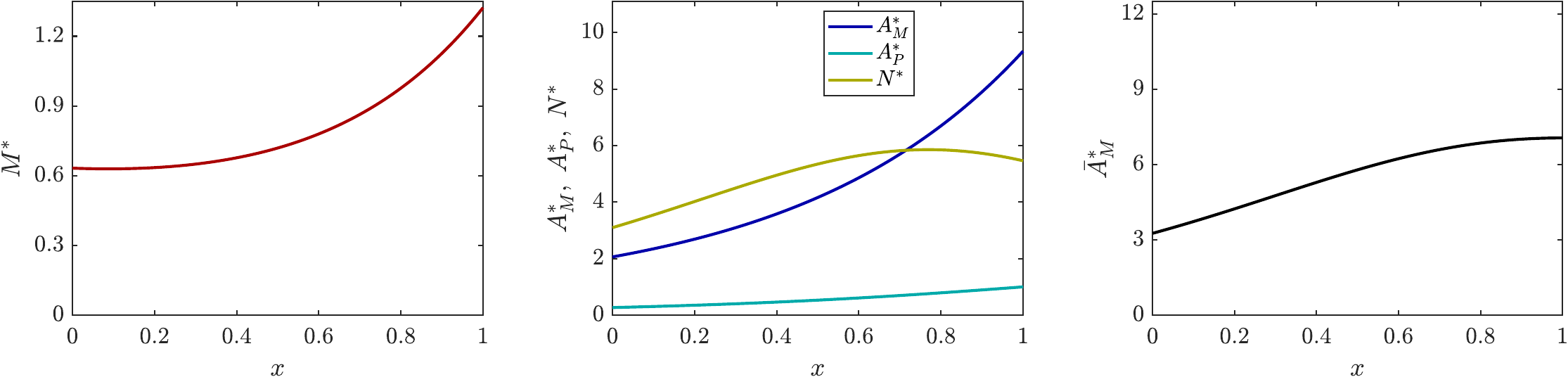}
		 \label{SS_plots_h15_s015}
		\vspace{-0.6cm}
	\end{subfigure}
	\par\medskip
	\begin{subfigure}[b]{\textwidth}
		\centering	
		\caption{$h=1.5,\,\sigma_M=0.3$}
		\includegraphics[height=3.96cm]{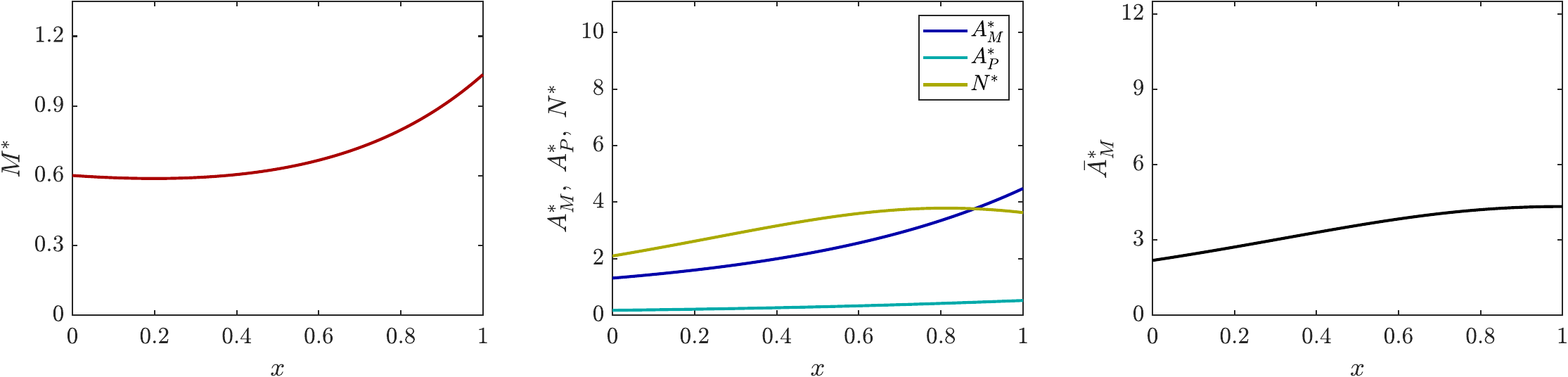}
		 \label{SS_plots_h15_s03}
		\vspace{-0.6cm}
	\end{subfigure}
	\par\medskip
	\begin{subfigure}[b]{\textwidth}
		\centering	
		\caption{$h=1.5,\,\sigma_M=1.5$}
		\includegraphics[height=3.96cm]{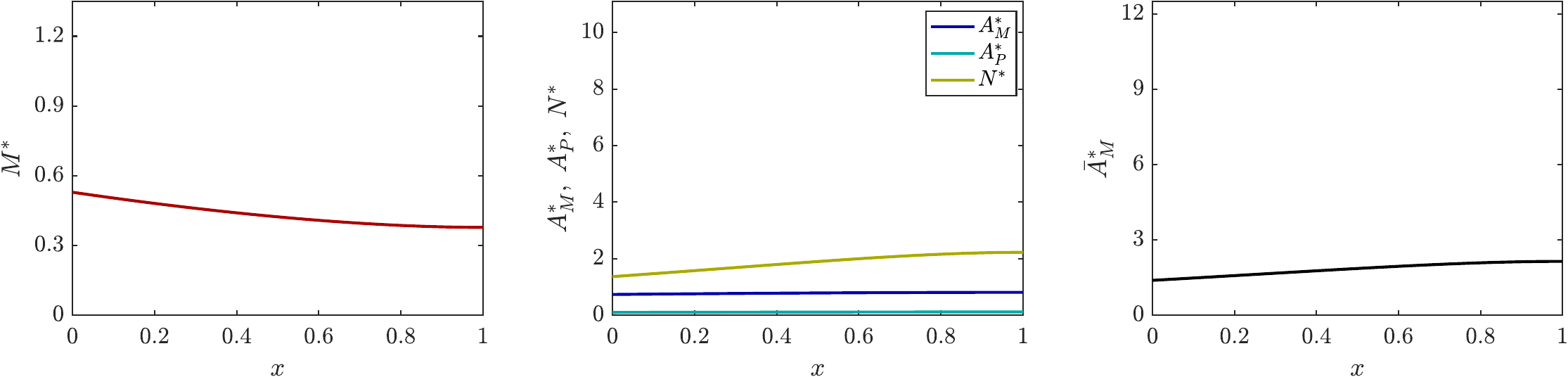}
		 \label{SS_plots_h15_s15}
		\vspace{-0.6cm}
	\end{subfigure}
	\par\medskip
	\begin{subfigure}[b]{\textwidth}
		\centering	
		\caption{$h=1.5,\,\sigma_M=3$}
		\includegraphics[height=3.96cm]{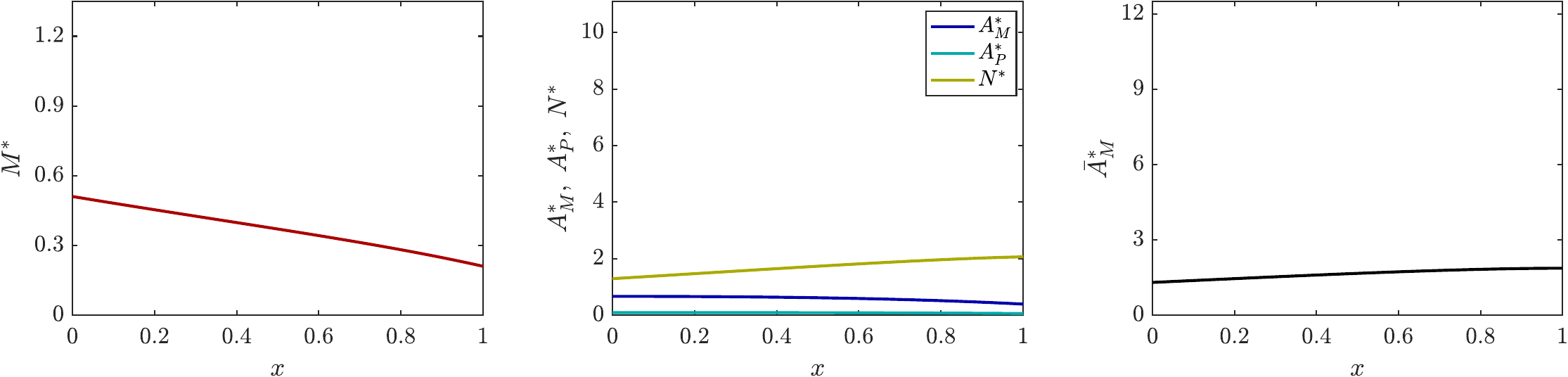}
		\label{SS_plots_h15_s30}
		\vspace{-0.6cm}
	\end{subfigure}
	\caption{Steady state solution profiles for $M^*$ (red lines; left), $A_M^*$ (blue lines; middle), $A_P^*$ (cyan lines; middle), $N^*$ (gold lines; middle), and $\bar{A}_M^*=A_M^*/M^*$ (black lines; right) for $h=1.5$ and (a) $\sigma_M=0.15$, (b) $\sigma_M=0.3$, (c) $\sigma_M=1.5$, or (d) $\sigma_M=3$. Expressions for $M^*$, $A_M^*$, $A_P^*$, and $N^*$ are given by equations (\ref{M_SS})--(\ref{N_SS}), respectively. All other parameters are set to the values given in Table~\ref{dimless_params}. From equation (\ref{res_time}), the approximate macrophage transit times are (a) $\tau\approx3.38$, (b) $\tau\approx1.87$, (c) $\tau\approx0.67$, and (d) $\tau\approx0.52$.} \label{SS_plots_h15}
\end{figure}

Representative examples of solution profiles obtained for a range of $\sigma_M$ values with a small chemotactic velocity ($h=0.3$) and a large chemotactic velocity ($h=1.5$) are shown in Figures~\ref{SS_plots_h03} and \ref{SS_plots_h15}, respectively. An interesting observation in these plots is that a seemingly minor change in the steady state cell profile $M^*$ can produce a substantial change in the amounts of intracellular, apoptotic and necrotic lipid held in the plaque (e.g., compare Figures \ref{SS_plots_h03_s015} and \ref{SS_plots_h03_s03}). This suggests a substantial change in the macrophage transit time between these two cases, which is confirmed by equation~(\ref{res_time}) (i.e., $\tau\approx6.11$ for Figure~\ref{SS_plots_h03_s015}, and $\tau\approx3.33$ for Figure~\ref{SS_plots_h03_s03}).

We begin our analysis by first highlighting two important properties of the model solutions:  
\begin{itemize}
\item The expressions for the lipid variables $A_M^*$, $A_P^*$ and $N^*$ are not well-defined when $s=\sigma_M=0$. This is because, when there is no flux of cells out of the domain, there is no mechanism for lipid removal from the system and the lipid variables blow up as $t \to \infty$. The cell variable $M^*$ does not experience blow up because the growth of the macrophage population is limited by apoptosis.
\item The average lipid per cell, defined by $\bar{A}_M^*\big(x\big) = \frac{A_M^*(x)}{M^*(x)}$, is always monotonic increasing. This is because the macrophages have a strictly positive average velocity and ingest lipid continuously. Consequently, using equation (\ref{AP_SS}) to rewrite equation (\ref{N_SS}) as:
\begin{equation}
	N^* \, =  \, \frac{\nu}{\theta} \cdot \frac{\bar{A}_M^*}{\nu \, + \, \eta M^*},
\end{equation}
we see that $N^*$ is monotonic increasing whenever $M^*$ is monotonic decreasing.
\end{itemize}

Further insight can be gained by a detailed analysis of the $M^*$ equation. For $x\in\mathbb{R}$, the expression on the right hand side of equation (\ref{M_SS}) is either monotonic decreasing or non-monotonic with a global minimum at some $x = x^*$. Whether this minimum exists, and its position relative to the domain boundaries, determines the qualitative profile of $M^*$ in the domain.

Considering $x^*$ as a function of $s$ for fixed $p$ and $r$ (i.e., effectively as a function of $\sigma_M$), an expression for the position of the minimum is:
\begin{equation}
x^*\big(s\big) \, = \, 1 \, - \, \frac{2}{\sqrt{p^2+4r}} \, \artanh\Bigg[{\frac{\big(s-p\big)\sqrt{p^2+4r}}{p\big(s-p\big)-2r}}\Bigg], \label{xstar_eq}
\end{equation} 
where the condition for existence of the minimum is:
\begin{equation}
	\frac{p-\sqrt{p^2+4r}}{2} \, < \, s \, < \, \frac{p+\sqrt{p^2+4r}}{2} \, .
\end{equation}
Notice that the lower inequality in the above expression is always satisfied for physical parameter values, so only the upper inequality is significant. Equation (\ref{xstar_eq}) shows that $x^*$ is an increasing function of $s$. Hence, when the minimum exists, its position moves up (down) the $x$-axis as the value of $s$ is increased (decreased). This has implications for the steady state profiles of the model variables, as summarised in the following three cases. 

\underline{\textbf{Case 1} ($s=p$ or $\sigma_M=h$)}:\newline
When $s=p$, the cell flux out of the domain at $x=1$ is exactly equal to the chemotactic flux. The boundary condition (\ref{SS_M_BC_1}) reduces to a condition of zero gradient (or zero diffusive flux). Consistent with this, equation (\ref{xstar_eq}) shows the existence of a minimum in $M^*$ at $x^*=1$. The $M^*$ profile is therefore monotonic decreasing for $x\in[0,1$), which implies that 
 $N^*$ is monotonic increasing on the same interval. Equation (\ref{AM_SS}) simplifies to:
\begin{equation}
	A_M^*\big(x\big) \, = \, \frac{r}{p} \, \big( 1 \, + \, \lambda \big) \, - \, \lambda  M^*\big(x\big) \, , \label{AM_SS_simp}	
\end{equation}
which further implies that $A_M^*$ and, hence, $A_P^*$ are monotonic increasing for $x\in[0,1$).

Examples of solution profiles with $s=p$ are shown in Figures~\ref{SS_plots_h03_s03} and \ref{SS_plots_h15_s15}. The case with the larger chemotactic velocity (Figure~\ref{SS_plots_h15_s15}) displays: (1) a smaller and more uniformly distributed macrophage population; (2) reduced overall quantities of intracellular, apoptotic and necrotic lipid; and (3) smaller average lipid loads per cell throughout the domain. All three of these observations can be attributed to the substantial 5-fold reduction in the average macrophage transit time from $\tau=\frac{10}{3}$ to $\tau=\frac{2}{3}$ (cf. equation~(\ref{res_time}) and Figure~\ref{transit_D_8_10}). 
 
\underline{\textbf{Case 2} ($s>p$ or $\sigma_M>h$)}:\newline
When $s>p$, the cell flux out of the domain at $x=1$ exceeds the chemotactic flux. The minimum described by equation (\ref{xstar_eq}) then either occurs outside the domain ($x^*>1$) or ceases to exist. Hence, $M^*$ is strictly monotonic decreasing and $N^*$ is strictly monotonic increasing for $x\in[0,1]$.

For $s>p$, the boundary condition (\ref{SS_AM_BC_1}) imposes a negative gradient in $A_M^*$ at $x=1$. The gradient in $A_M^*$ at $x=0$ is positive when the value of $s$ is close to the value of $p$. This creates a non-monotonic $A_M^*$ profile with a global maximum inside the domain (not shown). As the value of $s$ increases further beyond $p$, the gradient in $A_M^*$ at $x=0$ becomes negative and the $A_M^*$ profile is then monotonic decreasing (Figures \ref{SS_plots_h03_s06}, \ref{SS_plots_h03_s15} and \ref{SS_plots_h15_s30}). Observations indicate that the $A_P^*$ profile follows a similar pattern of behaviour. As the value of $s$ is increased beyond $p$, the $A_P^*$ profile first becomes non-monotonic with a global maximum inside the domain (Figure \ref{SS_plots_h03_s06}), and then becomes monotonic decreasing (Figure \ref{SS_plots_h03_s15}). When both $A_M^*$ and $A_P^*$ have a local maximum inside the domain, the local maximum in $A_P^*$ is observed to always occur at a larger $x$ than the local maximum in $A_M^*$.

\underline{\textbf{Case 3} ($s<p$ or $\sigma_M<h$)}:\newline
When $s<p$, the cell flux out of the domain at $x=1$ is smaller than the chemotactic flux and the cells experience a degree of trapping. The boundary condition (\ref{SS_M_BC_1}) imposes a positive gradient in $M^*$ at $x=1$, and, consequently, the $M^*$ 
profile in the domain must be either non-monotonic with a global minimum ($0 < x^* < 1$; Figures \ref{SS_plots_h03_s015}, \ref{SS_plots_h15_s015} and \ref{SS_plots_h15_s03}), or monotonic increasing ($x^* < 0$; not shown). As the value of $s$ is reduced downwards from $p$ to $0$ (for fixed $p$ and $r$), the transition from a non-monotonic to a monotonic increasing $M^*$ profile occurs at the critical value of $s$ for which $x^*=0$ in equation (\ref{xstar_eq}). However, as this critical value of $s$ can be negative for some $p$ and $r$, there exist cases where a non-monotonic $M^*$ profile is retained for all $0<s<p$, and a monotonic increasing $M^*$ profile can never be obtained.

\begin{figure}[h!]
	\centering		
	\includegraphics[height=6.6cm]{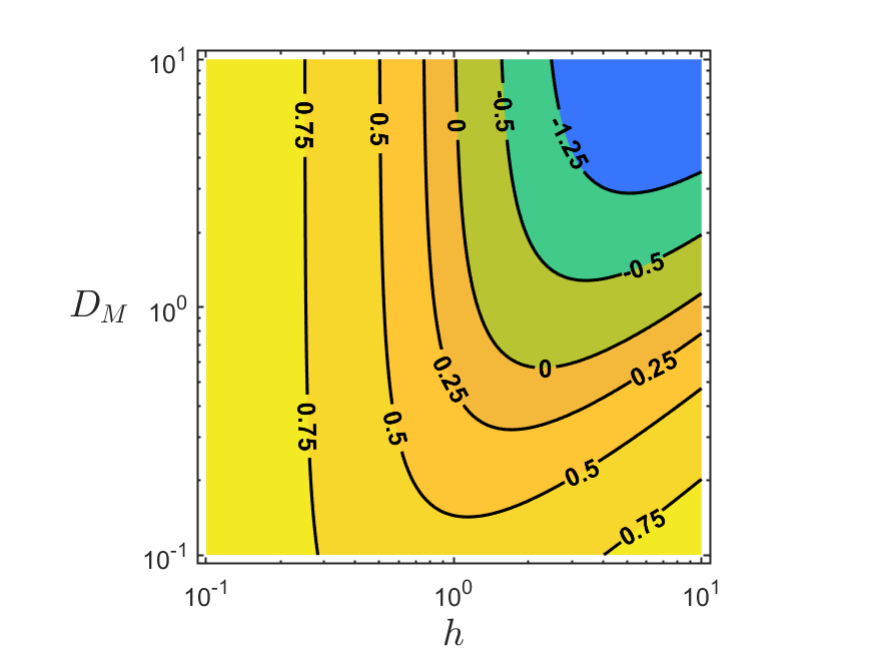}
	\caption{Contour plot showing the position $x^*<1$ of the global minimum of $M^*\left(x\right)$ for $\sigma_M=0$ and $\left(h,D_M\right)\in\left[0.1,10\right]\times\left[0.1,10\right]$. Expressions for $M^*\left(x\right)$ and $x^*$ are given by equations (\ref{M_SS}) and (\ref{xstar_eq}), respectively.}\label{xstar}
\end{figure}

Figure~\ref{xstar} presents a contour plot of $x^*$ as a function of $h$ and $D_M$ for the limiting case $s=\sigma_M=0$. The plot shows, for each combination of $h$ and $D_M$, the leftmost possible position of the global minimum $x^*$ for physical $\sigma_M$. Regions of the plot with $x^*>0$ (i.e., global minimum of $M^*$ inside the domain) produce non-monotonic $M^*$ profiles for all $0<s<p$. Regions of the plot with $x^*<0$ (i.e., global minimum of $M^*$ outside the domain) produce monotonic increasing $M^*$ profiles for $0<s<s_{crit}$, and non-monotonic $M^*$ profiles with a global minimum for $s_{crit}<s<p$, where $x^*\left(s_{crit}\right)=0$. The plot indicates that monotonic increasing $M^*$ profiles are only possible for $h$ and $D_M$ sufficiently large (with $s$, or $\sigma_M$, sufficiently small).

Observations suggest that, for $s<p$, the $A_M^*$ and $A_P^*$ profiles are both always monotonic increasing. The $N^*$ profile, however, becomes non-monotonic with a global maximum inside the domain (Figures \ref{SS_plots_h03_s015}, \ref{SS_plots_h15_s015} and \ref{SS_plots_h15_s03}). The position of the global maximum appears to move towards smaller $x$ with increasing extent of macrophage accumulation near $x=1$ (e.g., compare Figures \ref{SS_plots_h15_s015} and \ref{SS_plots_h15_s03}). This is because the rate of necrotic lipid consumption becomes larger in the deeper, rather than the central, regions of the plaque in this case.

Table \ref{SS_sol_summary} provides a summary of the possible qualitative solution profiles identified for each of the model variables for Cases 1, 2 and 3 above. We remark here that, under the assumption of passive rather than active macrophage emigration (i.e., $p=h=0$), only the steady state solution profiles described in Case 2 would be attainable.

\begin{table}[h!]
	\centering
	\footnotesize
	\renewcommand{\arraystretch}{1.5}
	\begin{tabular}{| c | >{\centering\arraybackslash}p{3.5cm} | >{\centering\arraybackslash}p{3.5cm} | >{\centering\arraybackslash}p{3.5cm} |}
		\hline
		\multirow{2}*{Variable} & \multicolumn{3}{ c |}{Spatial Solution Profile} \\ 
		\cline{2-4}
		& \textbf{Case 1} ($s=p$) & \textbf{Case 2} ($s>p$) & \textbf{Case 3} ($s<p$) \\ \hline
		$M^*\left(x\right)$ & Monotonic decreasing & Monotonic decreasing & Non-monotonic with global minimum OR monotonic increasing \\
		\hline
		$A_M^*\left(x\right)$ & Monotonic increasing & Non-monotonic with global maximum OR monotonic decreasing & Monotonic increasing \\
		\hline
		$A_P^*\left(x\right)$ & Monotonic increasing & Non-monotonic with global maximum OR monotonic decreasing & Monotonic increasing \\
		\hline
		$N^*\left(x\right)$ & Monotonic increasing & Monotonic increasing & Non-monotonic with global maximum \\
		\hline
	\end{tabular}
	\caption{Summary of the possible steady state solution profiles for each model variable as given by the equations (\ref{M_SS})--(\ref{N_SS}).} \label{SS_sol_summary}
\end{table}
 
\subsection{Full Model (Spatially Non-Uniform Chemotaxis)}\label{results_full}
\begin{figure}[h!]
	\centering		
	\includegraphics[height=7cm]{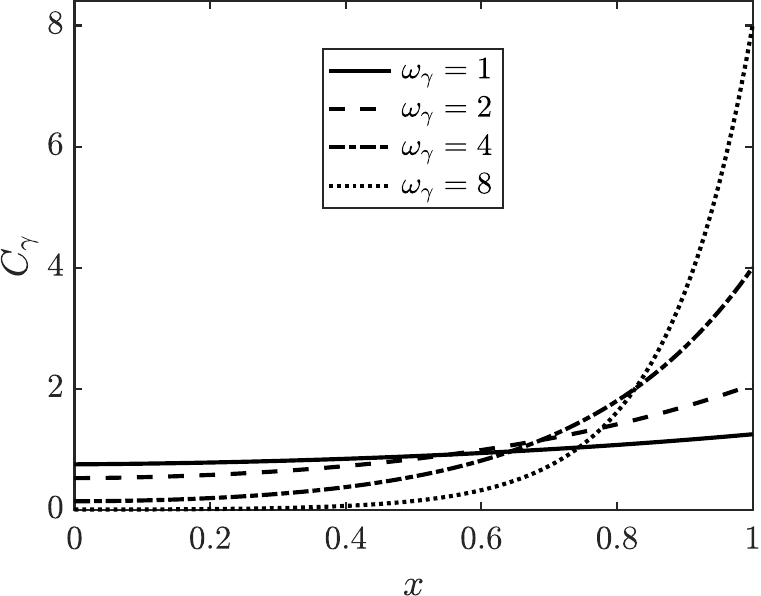}
	\caption{Spatial profiles of the dimensionless chemoattractant concentration $C_\gamma\left(x\right)$ given by equation (\ref{dimless_C_eqn}) for parameter values $\sigma_\gamma=0.1$ and $\omega_\gamma=1$ (solid line), $\omega_\gamma=2$ (dashed line), $\omega_\gamma=4$ (dash-dot line), or $\omega_\gamma=8$ (dotted line).
	}\label{C_gamma}
\end{figure}

In this section, we relax the assumption of spatially uniform chemotaxis and consider the full model. Thus, the macrophage chemotactic velocity now varies spatially according to the chemoattractant profile as determined by the values of the parameters $\omega_\gamma$ and $\sigma_\gamma$. Throughout this section, we fix $\sigma_\gamma=0.1$ and assume $\omega_\gamma \geqslant 1$ so that the chemical diffusion distance in the intima is comparable to, or smaller than, the length of the domain. Examples of dimensionless chemoattractant profiles for $\omega_\gamma=1,2,4,8$ are shown in Figure~\ref{C_gamma}. As macrophages traverse the domain, their dimensionless chemotactic velocity $\chi_\gamma\,\frac{dC_\gamma}{dx}$ now increases monotonically from a minimum value of $\chi_\gamma \omega^2_\gamma \cdot \big[\frac{\omega_\gamma}{\sigma_\gamma}\sinh(\omega_\gamma) + \cosh(\omega_\gamma)\big]^{-1}$ at $x=0$ to a maximum value of $\chi_\gamma\,\omega_\gamma^2$ at $x=1$.

We assume that the dimensional value of the chemical diffusion coefficient $D_\gamma$ is fixed and consider the dimensionless $\omega_\gamma$ value to be determined by the dimensional decay rate $\mu_\gamma$. A $n$-fold change in $\omega_\gamma$ is therefore interpreted as a $n^2$-fold change in $\mu_\gamma$ (see Table \ref{dimless_params}), which alters both the scaling of the dimensional chemoattractant concentration (see equation~(\ref{nondim})) and the value of the dimensionless chemotaxis coefficient $\chi_\gamma$ (see Table \ref{dimless_params}).

To facilitate comparison with the results for fixed chemotactic velocity $h$ in Section~\ref{results_reduced}, here we define the spatially averaged chemotactic velocity:
\begin{equation}
	\bar{h}=\int_0^1 \chi_\gamma \, \frac{dC_\gamma}{dx} \, dx \, = \, \chi_\gamma \, \omega_\gamma \, \Bigg[\frac{\omega_\gamma \, \big(\cosh(\omega_\gamma)-1\big)+\sigma_\gamma \, \sinh(\omega_\gamma)}{\omega_\gamma \, \sinh(\omega_\gamma)+\sigma_\gamma \, \cosh(\omega_\gamma)}\Bigg] \, .
\end{equation}
In practice, we use this definition to calculate the value of the chemotaxis coefficient $\chi_\gamma$ that produces a desired value of $\bar{h}$ for a given $\omega_\gamma$.

\begin{figure}[h!]
	\centering
	\begin{subfigure}[b]{0.49\textwidth}
		\centering	
		\includegraphics[height=5.2cm]{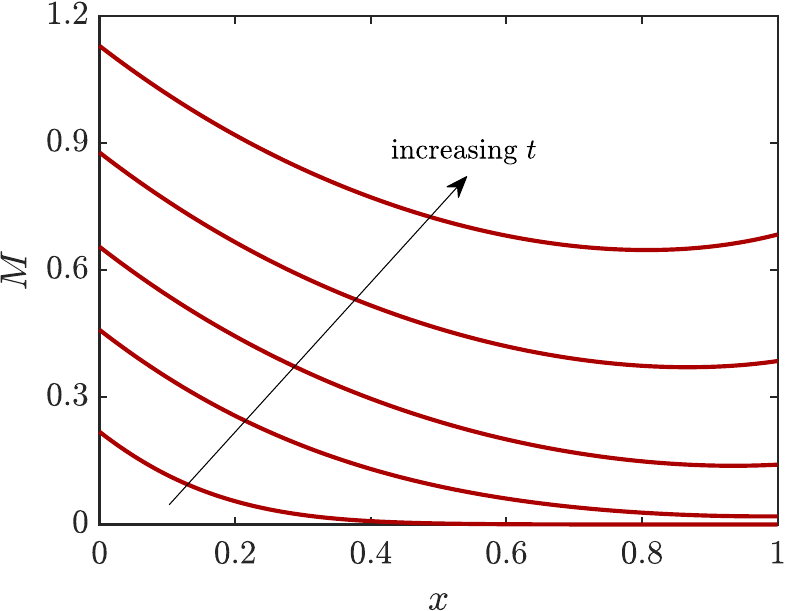}
		\caption{} \label{transient_M}
	\end{subfigure}
	\begin{subfigure}[b]{0.49\textwidth}
		\centering	
		\includegraphics[height=5.2cm]{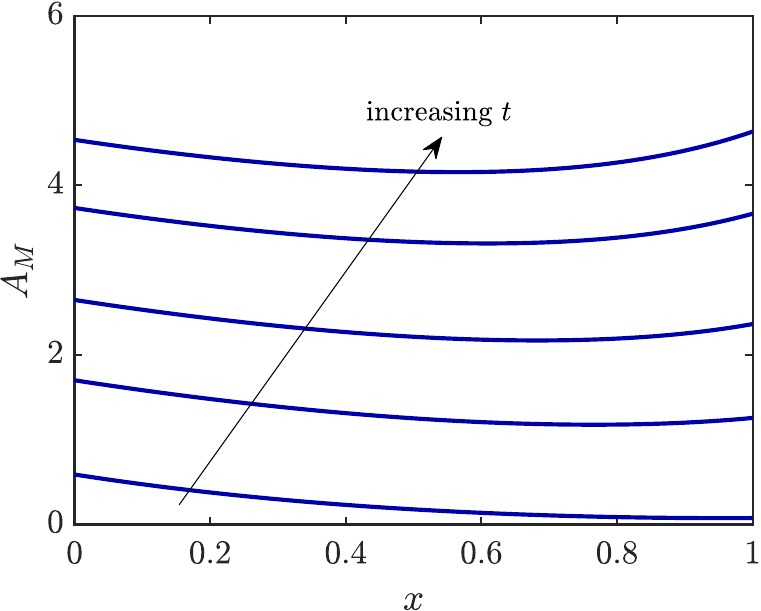}
		\caption{} \label{transient_AM}
	\end{subfigure}
	\par \medskip
	\begin{subfigure}[b]{0.49\textwidth}
		\centering	
		\includegraphics[height=5.2cm]{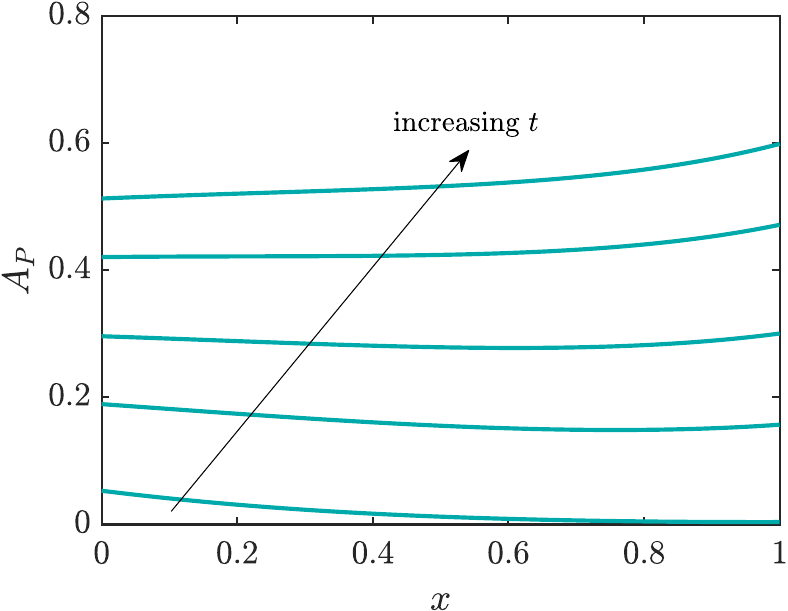}
		\caption{} \label{transient_AP}
	\end{subfigure}
	\begin{subfigure}[b]{0.49\textwidth}
		\centering	
		\includegraphics[height=5.2cm]{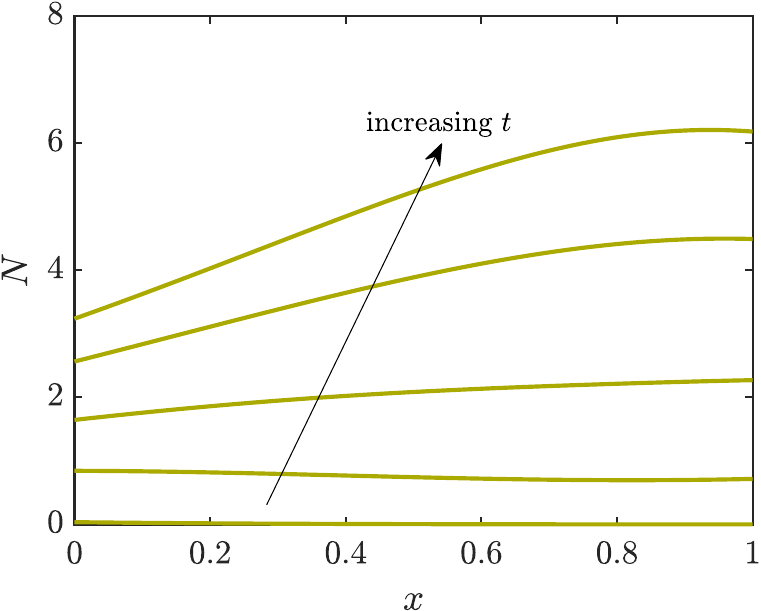}
		\caption{} \label{transient_N}
	\end{subfigure}
	\caption{Full model solution profiles for (a) $M$, (b) $A_M$, (c) $A_P$, and (d) $N$ at several times $t$. Solutions for $M$ are plotted at the (approximate) times $t=0.03$, 0.15, 0.35, 0.8 and 3.2. Solutions for $A_M$, $A_P$ and $N$ are plotted at the times $t=0.24$, 2.1, 5.0, 11.5 and 43.6. Directions of increasing time are indicated by arrows. Each profile is near steady state at the final $t$. Solutions use the parameter values $\omega_\gamma = 2$, $\bar{h} = 0.3$ ($\chi_\gamma \approx 0.1944$), and $\sigma_M = 0.3$. All other parameters are set to the values given in Table~\ref{dimless_params}.} \label{transient}
\end{figure}

Transient solutions from a numerical simulation of the full model with $\omega_\gamma = 2$, $\bar{h} = 0.3$ ($\chi_\gamma \approx 0.1944$), and $\sigma_M = 0.3$ are shown in Figure~\ref{transient}. The spatial profile of each variable is plotted at several times, where the final time represents an approximate steady state. For the cell variable $M$ (Figure~\ref{transient_M}), a steady profile emerges after 2 to 3 dimensionless time units (around 2 weeks of physical time). For the lipid variables $A_M$, $A_P$ and $N$ (Figures~\ref{transient_AM}, \ref{transient_AP} and \ref{transient_N}, respectively), a steady profile emerges after more than 40 dimensionless time units (around 7 months of physical time). This implies that, beyond the initial phase of cell accumulation, there is a steady influx of new macrophages, each generation of which has cells that acquire increasingly large lipid loads before eventually dying or leaving the plaque by emigration.

\begin{figure}[h!]
	\centering
	\begin{subfigure}[b]{0.49\textwidth}
		\centering	
		\includegraphics[height=5.2cm]{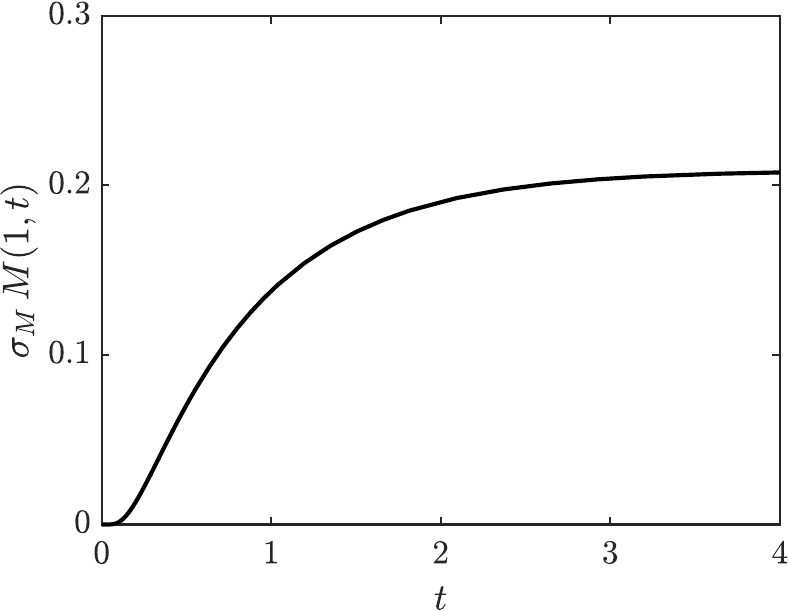}
		\caption{} \label{efflux_M}
	\end{subfigure}
	\begin{subfigure}[b]{0.49\textwidth}
		\centering	
		\includegraphics[height=5.2cm]{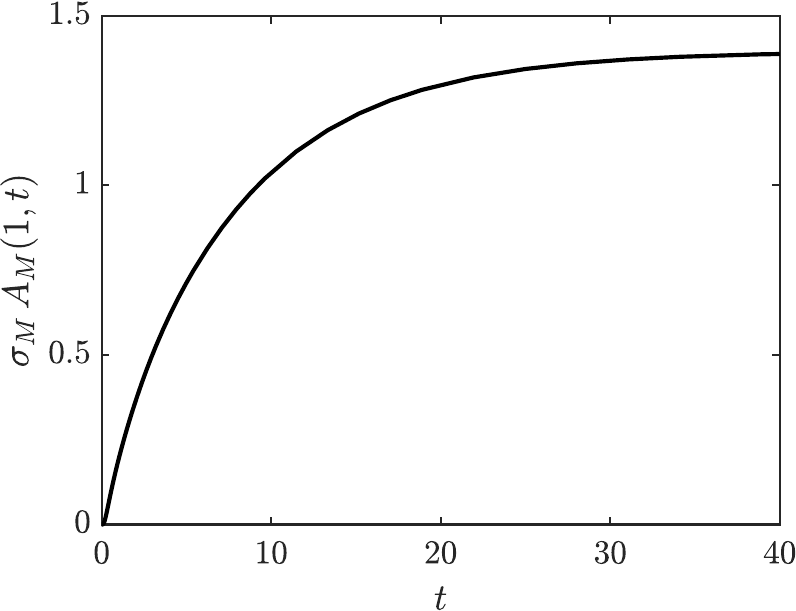}
		\caption{} \label{efflux_AM}
	\end{subfigure}
	\caption{Plots of (a) the macrophage flux $\sigma_M M(1,t)$ and (b) the internalised lipid flux $\sigma_M A_M(1,t)$ at the IEL against time $t$ for the full model with $\sigma_M=0.3$, $\omega_\gamma=2$, and $\bar{h}=0.3$ ($\chi_\gamma \approx 0.1944$). All other parameters are set to the values given in Table~\ref{dimless_params}. Note the difference in timescales between the two plots.} \label{efflux}
\end{figure}

Figure~\ref{efflux} quantifies the role of emigration in this simulation by plotting the cell flux $\sigma_M M(1,t)$ (Figure~\ref{efflux_M}) and the internalised lipid flux $\sigma_M A_M(1,t)$ (Figure~\ref{efflux_AM}) at the IEL over time. These plots show that, in the long term, only 20\% of the cells that enter the plaque ultimately emigrate, while the lipid load removed from the plaque by each cell eventually approaches an average of around 7 endogenous lipid units.

\begin{figure}[h!]
	\centering
	\begin{subfigure}[b]{\textwidth}
		\centering	
		\includegraphics[height=3.96cm]{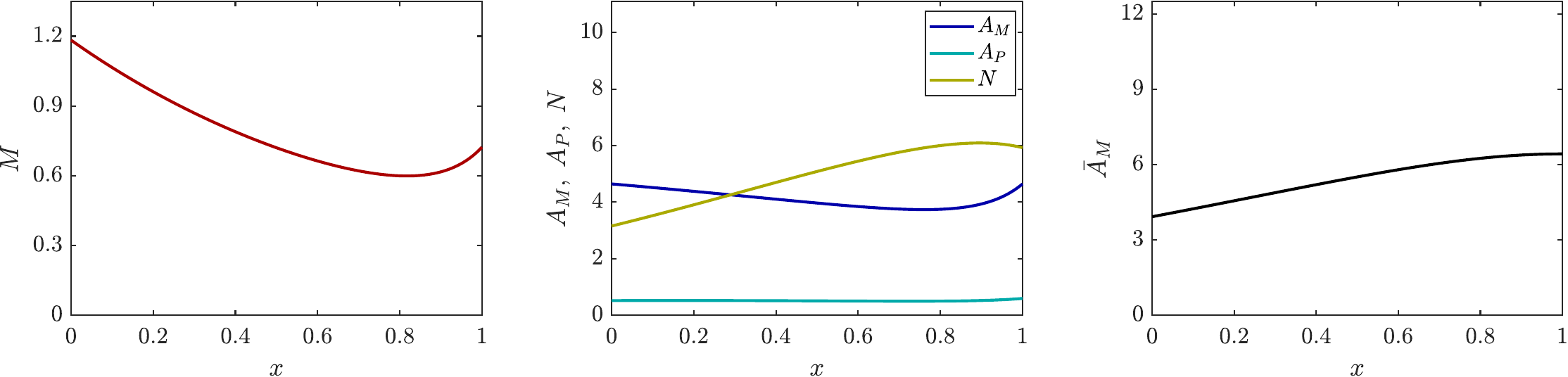}
		\caption{$\bar{h}=0.3,\,\omega_\gamma=8$} \label{SS_plots_full_om8}
	\end{subfigure}
	\par\medskip
	\begin{subfigure}[b]{\textwidth}
		\centering	
		\includegraphics[height=3.96cm]{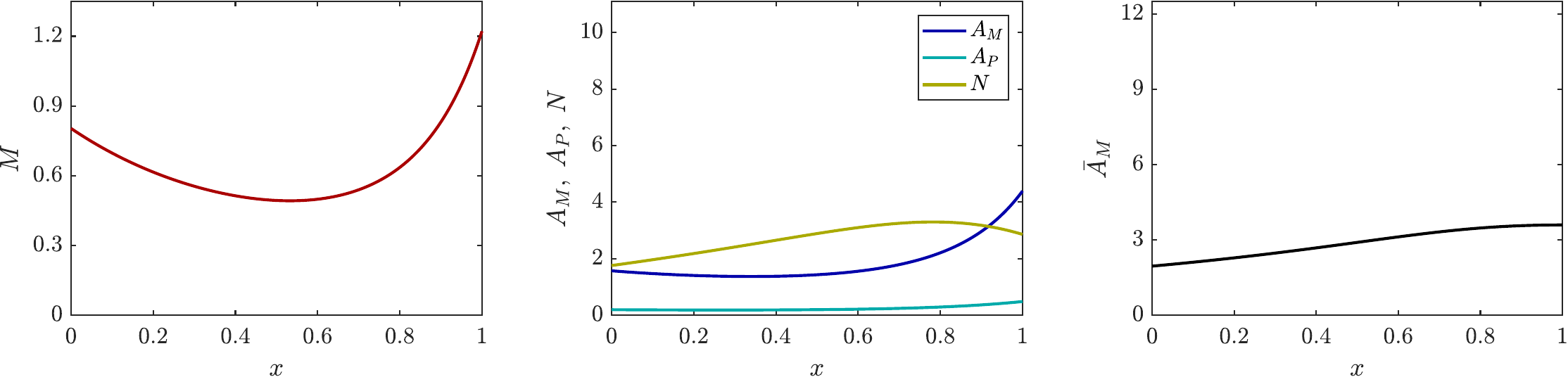}
		\caption{$\bar{h}=1.5,\,\omega_\gamma=2$} \label{SS_plots_full_hbar15}
	\end{subfigure}
	\caption{Long-time full model solution profiles for $M$ (red lines; left), $A_M$ (blue lines; middle), $A_P$ (cyan lines; middle), $N$ (gold lines; middle), and $\bar{A}_M=A_M/M$ (black lines; right) for $\sigma_M=0.3$ and (a) $\bar{h}=0.3$, $\omega_\gamma=8$ ($\chi_\gamma \approx 0.0375$), or (b) $\bar{h}=1.5$, $\omega_\gamma=2$ ($\chi_\gamma \approx 0.9720$). All other parameters are set to the values given in Table~\ref{dimless_params}.} \label{SS_plots_full}
\end{figure}

Comparison of Figure~\ref{transient} with Figure~\ref{SS_plots_h03_s03} shows that, for these particular parameter values, the long-time numerical solutions of the full model are quantitatively similar to the steady state solutions of the reduced model. To further explore differences between outcomes for the full model and the reduced model, Figure~\ref{SS_plots_full} presents long-time numerical solutions for two different modifications to the scenario simulated in Figure~\ref{transient}. We first consider $\omega_\gamma = 8$ so that the chemical concentration gradient has greater spatial variation (Figure~\ref{SS_plots_full_om8}), and we then consider $\bar{h} = 1.5$ so that the overall macrophage chemotactic response is stronger (Figure~\ref{SS_plots_full_hbar15}). The plots in Figures~\ref{SS_plots_full_om8} and \ref{SS_plots_full_hbar15} are suitable for comparison against the steady state solutions of the reduced model given in Figures~\ref{SS_plots_h03_s03} and \ref{SS_plots_h15_s03}, respectively.

In the case with $\omega_\gamma=8$ ($\chi_\gamma=0.0375$), the chemoattract penetrates only a small distance into the domain (see Figure~\ref{C_gamma}). Macrophage chemotaxis is therefore predominant only in the vicinity of the IEL. Compared to Figure~\ref{SS_plots_h03_s03} for the reduced model, we see an increase in the steady state cell density at both ends of the domain (Figure~\ref{SS_plots_full_om8}; left panel). At $x=0$, this is because the absence of chemotaxis slows the movement of cells away from the boundary, while at $x=1$, this is because the local chemotactic velocity ($\chi_\gamma\,\omega_\gamma^2 \approx 2.4$) exceeds the IEL permeability ($\sigma_\gamma=0.3$) and we are in the ``trapping'' regime (\textbf{Case 3}) discussed in Section~\ref{results_reduced_SS}. Compared to Figure~\ref{SS_plots_h03_s03}, the non-uniform chemotaxis also leads to a slight reduction in the lipid densities $A_M$, $A_P$, and $N$ across the domain (Figure~\ref{SS_plots_full_om8}; middle panel). Overall, however, the effects of the non-uniform chemotaxis are quantitatively small in this case, which suggests that the reduced model can provide a reasonable quantitative approximation to the long-term dynamics of the full model for a wide range of $\omega_\gamma$ values provided that the overall chemotactic response, as quantified by $\bar{h}$, is relatively weak.

In the case with $\bar{h}=1.5$ ($\chi_\gamma=0.9720$), we observe that the long-time solutions of the full model are substantially different to those for the reduced model despite the relatively gentle spatial variation in the chemoattractant gradient ($\omega_\gamma=2$). The cell profile, in particular, shows a large qualitative change due to the pronounced elevation of cell density at both ends of the domain by the same mechanisms as described above (Figure~\ref{SS_plots_full_hbar15}; left panel). The feedthrough effect of the spatially non-uniform macrophage chemotaxis is a noticeable reduction in the steady lipid densities $A_M$, $A_P$, and $N$ across the domain (Figure~\ref{SS_plots_full_hbar15}; middle panel), which is perhaps best exemplified by the reduction in average lipid load $\bar{A}_M$ for all $x$ (right panel). These results suggest that, if we control for average chemotactic velocity $\bar{h}$, the overall lipid burden in the model plaque can be reduced if directed macrophage migration is constrained to occur primarily in a region local to the IEL.

\begin{figure}[h!]
	\centering		
	\includegraphics[height=6.6cm]{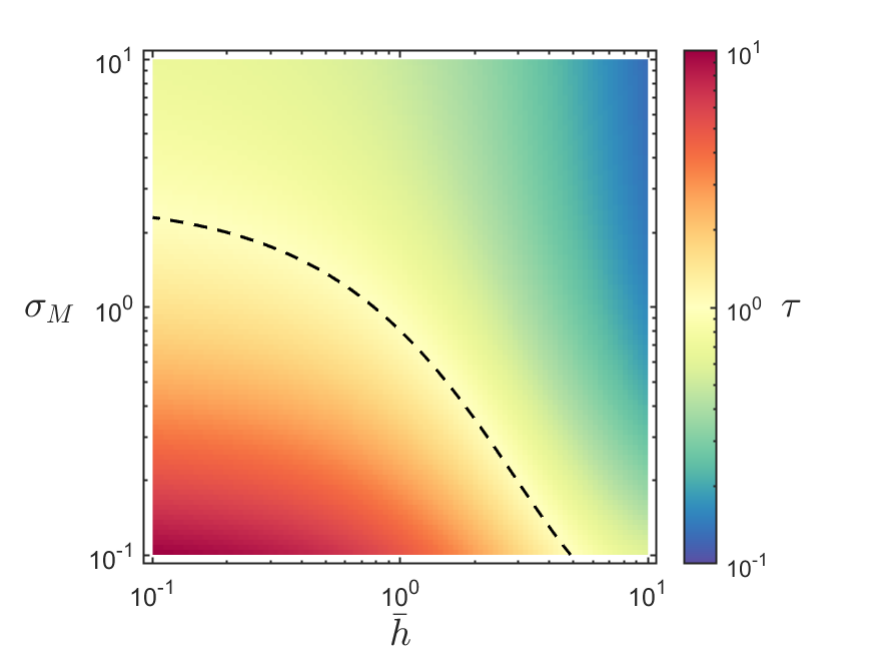}
	\caption{Heatmap of average macrophage transit time $\tau$ calculated for equation~(\ref{SS_M_trans_full}) with boundary conditions (\ref{dimless_M_BC_0}) and (\ref{dimless_M_BC_1}) for $\left(\bar{h},\sigma_M\right)\in\left[0.1,10\right]\times\left[0.1,10\right]$ with $D_M=0.8$, $\sigma_\gamma=0.1$, and $\omega_\gamma=2$ ($\chi_\gamma\approx0.6480\,\bar{h}$). The dashed black line indicates the contour $\tau=1$.}\label{transit_full}
\end{figure}

The potential for spatially non-uniform chemotaxis to alter both the macrophage distribution and the overall lipid burden in the model plaque suggests that the characteristics of the spatial chemoattractant profile may have an interesting influence on macrophage transit times. Figure~\ref{transit_full} presents the transit time distribution as a function of $\bar{h}$ and $\sigma_M$ with $D_M=0.8$ and a chemoattractant profile given by $\sigma_\gamma=0.1$ and $\omega_\gamma=2$. As a closed form analytical solution cannot be obtained for the steady state equation:
\begin{equation}
	D_M \frac{d^2 M^*}{dx^2} \, - \, \chi_\gamma \, \frac{\partial}{\partial x} \, \Big( \, M^* \, \frac{dC_\gamma}{dx} \, \Big) \, = \, 0 \, , \label{SS_M_trans_full} \\
\end{equation}
with boundary conditions (\ref{dimless_M_BC_0}) and (\ref{dimless_M_BC_1}), transit times $\tau$ are instead generated numerically by solving the time-dependent equation to steady state. Comparison of Figure~\ref{transit_full} with Figure~\ref{transit_D_8_10} for the uniform chemotaxis model shows a clear leftward shift in the contour $\tau=1$ (dashed lines), which indicates that non-uniform chemotaxis can substantially reduce transit times. This is consistent with the observations of reduced long-term lipid burden for the simulations presented in Figure~\ref{SS_plots_full}. 

\begin{figure}[h!]
	\centering
	\begin{subfigure}[b]{0.49\textwidth}
		\centering	
		\includegraphics[height=6.6cm]{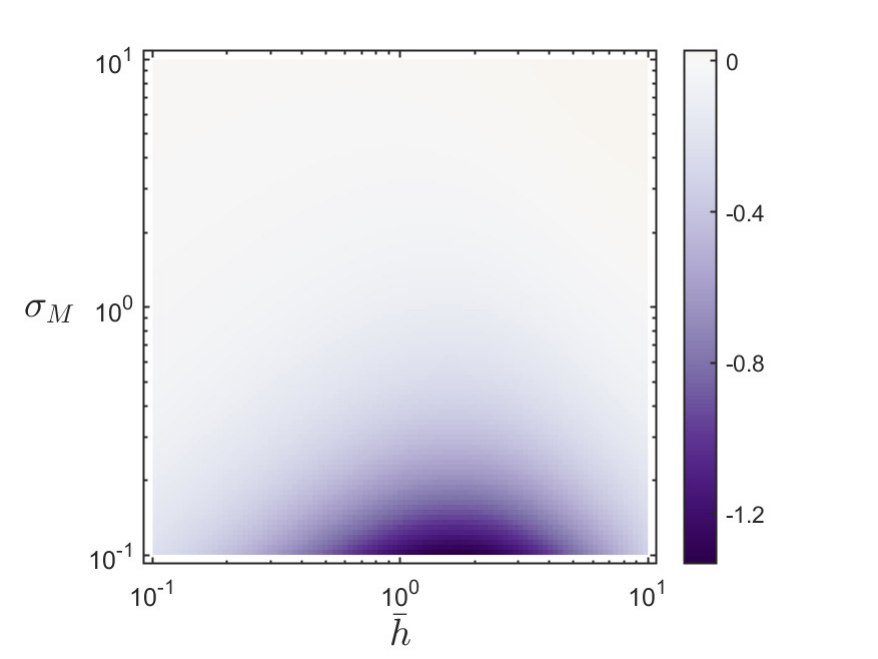}
		\caption{} \label{transit_compare_abs}
	\end{subfigure}
	\begin{subfigure}[b]{0.49\textwidth}
		\centering	
		\includegraphics[height=6.6cm]{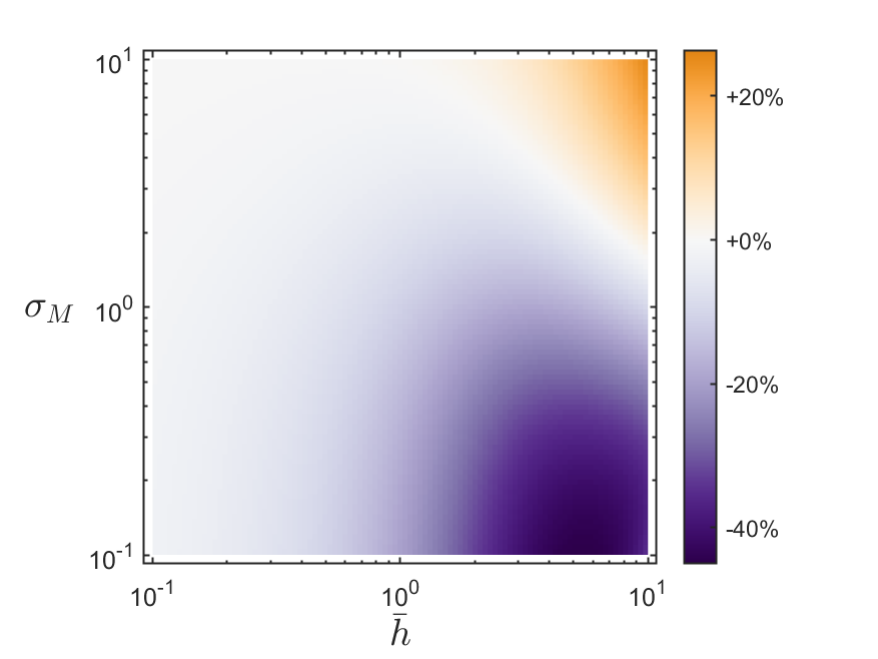}
		\caption{} \label{transit_compare_rel}
	\end{subfigure}
	\caption{Heatmaps of the (a) absolute change and (b) relative change in the transit time distribution for the spatially non-uniform chemotaxis model (Figure~\ref{transit_full}) versus the spatially uniform chemotaxis model (Figure~\ref{transit_D_8_10}). Negative values indicate reduced transit times in the spatially non-uniform model.
	} \label{transit_compare}
\end{figure}

Figure~\ref{transit_compare} presents the absolute change (Figure~\ref{transit_compare_abs}) and the relative change (Figure~\ref{transit_compare_rel}) in the transit time distribution for the full model (Figure~\ref{transit_full}) versus the reduced model (Figure~\ref{transit_D_8_10}). These plots show that, for $\omega_\gamma=2$, the spatially non-uniform chemotaxis model leads to reduced macrophage transit times for a wide range of $\bar{h}$ and $\sigma_M$ values. The absolute change in transit times is most significant for intermediate $\bar{h}$ (e.g., $0.4<\bar{h}<4$) and small $\sigma_M$ (e.g., $\sigma_M<0.5$).

\section{Discussion}
Chemotactically-guided macrophage emigration is a potentially effective mechanism for the removal of inflammatory lipid species from atherosclerotic plaques that has been implicated in studies of plaque regression \citep{Feig10,Feig11,vanG12}. However, the mechanisms that regulate this process \emph{in vivo} are not well understood, and it is thought that factors such as inflammation may inhibit the macrophage emigratory response. Given this uncertainty in the biology, this study proposes a model to address the general question: if chemotactic emigration of plaque macrophages occurs to some degree, what are the implications for the spatio-temporal dynamics of lipids and cells in the early plaque? We discuss our findings in the following sections.

\subsection{Macrophage Transit Time Indicates Emigratory Capacity and Early Plaque Pathology}
Our results indicate that it is beneficial to consider the implications of chemotactic emigration through the lens of the macrophage transit time $\tau$ (i.e., the average time required for a newly-recruited macrophage to traverse and then exit the plaque by a combination of random and directed movement). Using a reduced version of the model with spatially uniform chemotaxis, we derive the analytical expression (\ref{res_time}) for $\tau$ as a function of the dimensionless diffusion coefficient $D_M$, the (constant) chemotactic velocity $h=\chi_\gamma \, \frac{dC_\gamma}{dx}$, and the IEL permeability $\sigma_M$. A critical consideration in the interpretation of this expression is the value of $\tau$ relative to the average macrophage lifespan (which is scaled to 1 in dimensionless terms). Naturally, we find that $\tau$ decreases with increasing $h$ meaning that an increasing proportion of macrophages will emigrate out of the plaque rather than die \emph{in situ}. However, even if $h$ is relatively large, we find that a macrophage may remain more likely to die in the plaque than to emigrate if either $D_M$ or $\sigma_M$ is too low. This observation for $\sigma_M$ emphasises that the capacity for macrophages to effectively traverse the IEL is an important determinant of the overall effectiveness of the emigratory response. We discuss the role of the IEL as a physical barrier to macrophage emigration in more detail below.

We find that the macrophage transit time $\tau$ can also be used to understand the extent of model plaque lipid accumulation. In the model, the relative balance between macrophage apoptosis and emigration informs the total lipid quantity that ultimately accumulates in the $A_M$, $A_P$, and $N$ compartments. A model plaque dominated by apoptosis will accumulate a substantial amount of lipid, whereas a model plaque dominated by emigration will accumulate only a small amount of lipid. As the balance between emigration and apoptosis can be inferred by the value of $\tau$, we find that $\tau$ provides a reliable indicator of how the long-term lipid burden in the model plaque, and thus the early plaque pathology, varies with the macrophage transport parameters.

By controlling for the spatially-averaged macrophage chemotactic velocity $\bar{h}=\int_0^1 \chi_\gamma \, \frac{dC_\gamma}{dx} \, dx$, numerical simulations show that the qualitative features of the transit time distribution are preserved as the spatial profile of the emigratory chemoattractant is varied. However, we do observe noticable quantitative differences between predicted transit times for certain parameter combinations. For example, for moderate $\bar{h}$ and small $\sigma_M$, the transit time in the full model with $\omega_\gamma=2$ is reduced by more than 1 time unit (approximately 1 week of physical time) compared to the reduced model ($\omega_\gamma \ll 1$). Such reductions, which appear to persist over a range of $D_M$ and $\omega_\gamma$ values, presumably reflect a net increase in the overall spatially-averaged macrophage velocity that accounts for the role of diffusion. Overall, this observation suggests that increased localisation of the chemoattractant signal towards the IEL can lead to an increased proportion of macrophages that emigrate out of the plaque. This, of course, relies upon a sufficient rate of random motion to ensure that macrophages reach the signal on a timescale shorter than the typical cell lifespan.

\subsection{Ease of Macrophage Passage Through the IEL Determines the Spatial Characteristics of the Model Plaque}
The steady state solution of the reduced model presented in Section~\ref{results_reduced_SS} provides insight into the spatial distribution of the model cell and lipid species in the presence of emigratory chemotaxis. We identify all possible qualitative steady state solution profiles for each variable, and show that there exist three different regimes of solution profiles depending on the relative values of the dimensionless parameters $p=\frac{h}{D_M}$ and $s=\frac{\sigma_M}{D_M}$ (see Table~\ref{SS_sol_summary}). For fixed $D_M$, these regimes are determined by the relative values of the dimensionless macrophage chemotactic velocity $h$, and the dimensionless net velocity $\sigma_M$ at which emigrating macrophages pass through the IEL.

In light of this observation, it is natural to consider which of these parameter regimes is most realistic in practice. Given that fenestrae in the IEL may only be several microns in diameter \citep{Sand09}, we anticipate that the passage of macrophages through these holes may be slow and heavily reliant upon active (chemotactic) rather than passive (diffusive) migration. This suggests that the condition $s\leqslant p$ (\textbf{Case 1} or \textbf{Case 3}) may be more realistic than the condition $s>p$ (\textbf{Case 2}), in which case the model predicts the possibility of elevated macrophage density deep in the plaque due to the rate-limiting effect of the IEL on macrophage emigration. However, we note that the IEL can undergo dynamic remodelling during plaque progression, potentially leading to enlarged fenestral holes that increase the IEL permeability to cells \citep{Kwon98}.

An interesting observation of the model is that, across all possible parameter regimes, the variable that demonstrates the least variability in its qualitative steady state solution profile is the necrotic lipid density $N$. The necrotic lipid profile either increases monotonically across the intima ($s\geqslant p$), or has a local maximum inside the intima ($s<p$). For the parameter values considered in this study, any observed local maximum is typically shallow and tends to occur in the vicinity of the IEL (i.e., $x>0.7$). Hence, taking, for example, the centre of mass $X$ of the necrotic lipid distribution as a broad measure of necrotic core localisation, the model predicts that the early necrotic core will always localise deep in the plaque (i.e., $X>0.5$).

\subsection{Model Implications in the Context of Murine Experiments}
The transient model solutions presented in Figure~\ref{transient} demonstrate that the cell and lipid variables in the model evolve on two distinct timescales. The macrophage dyamics occur on a timescale of several weeks, whereas the lipid dynamics occur on a timescale of several months. Such dynamics have previously been observed in an ODE model of plaque formation, motivating a two-timescale analysis of the system \citep{Lui21}. In the context of murine plaque formation, the observed separation of timescales may be signficant. It has been reported that smooth muscle cells can enter plaques within 5 to 6 weeks of plaque initiation \citep{Misr18}, and a recent model of plaque SMC accumulation predicts that this can fundamentally alter the subsequent plaque dynamics \citep{Nden25}. This suggests that the current model may only be applicable on a timescale of 1 to 2 months, so that the predicted steady solutions for the lipid variables may never be attained in practice.

The model presented in this paper could potentially be used in conjunction with experimental data to help identify evidence of macrophage emigration during the first several weeks of plaque formation. Unfortunately, however, it seems rare for murine plaque data to be collected during early plaque development prior to the entry of smooth muscle cells. Given suitable data, such as the spatial distribution of macrophage density through an intimal cross-section, the contribution (or otherwise) of emigration could be inferred by comparison to the qualitative macrophage profiles identified here. Somewhat counter-intuitively, the model suggests that evidence of macrophage accumulation proximal to the IEL could be a hallmark of macrophage emigration because it would imply that the cell motility includes a bias towards the deep plaque.

\subsection{Limitations of the Model}
The above discussion should, of course, be considered in the context of the limitations of the model. The proposed model is intentionally parsimonious, aiming to allow for both detailed mathematical analysis and a clear focus on understanding the implications of active macrophage emigration. In particular, the model does not explicitly account for the role of HDL, and suppresses the details of LDL transport and retention by assuming a uniformly-distributed and inexhaustable source of tissue-bound lipid available for macrophage consumption. However, given that we consider the model to represent a plaque in the narrow murine intima, and that typical mouse experiments induce plaque formation by aggressive feeding of a high-fat diet, we believe these assumptions to be reasonable.

Other limitations of the model could potentially be addressed by building the assumed mechanisms of chemotactic emigration into existing spatial models of the early plaque. For example, the impact of tissue crowding and dynamic expansion of the intima could be studied in the multiphase PDE framework developed by \citet{Ahme23}. Moreover, the impact of lipid loading on macrophage motility, which has been proposed to inhibit macrophage emigration \citep{Moor18,Kang21}, could be studied by applying theory developed in the spatially resolved and lipid-structured model of \citet{Cham25}. 

\section{Conclusions}
This study presents a novel model of early atherosclerotic plaque formation that uses chemotaxis to capture the biological mechanisms of macrophage emigration more accurately than previous studies. The elegant framework allows for detailed mathematical analysis that provides insight into the implications of chemotactic macrophage emigration for the pathology and spatial characteristics of the early murine plaque. The theory developed herein can potentially be built into more detailed models of plaque formation, and provides an early step towards the development of models to better understand the dynamics of plaque regression.

\section*{Acknowledgements}
The author thanks Keith Chambers and Helen Byrne (Mathematical Institute, University of Oxford), and Mary Myerscough (School of Mathematics and Statistics, University of Sydney) for helpful discussions.

\bibliographystyle{plainnat}

\end{document}